\documentclass{article}

\usepackage[english]{babel}

\usepackage[letterpaper,top=2cm,bottom=2cm,left=3cm,right=3cm,marginparwidth=1.75cm]{geometry}

\usepackage{amsmath}
\usepackage{graphicx}% http://ctan.org/pkg/graphicx
\usepackage{array}% http://ctan.org/pkg/array
\usepackage{longtable}
\usepackage{booktabs}
\usepackage[colorlinks=true, allcolors=blue]{hyperref}
\usepackage{multirow}

\usepackage{color,soul}

\usepackage{scalerel}
\usepackage{tikz}
\usetikzlibrary{svg.path}

\definecolor{orcidlogocol}{HTML}{A6CE39}
\tikzset{
  orcidlogo/.pic={
    \fill[orcidlogocol] svg{M256,128c0,70.7-57.3,128-128,128C57.3,256,0,198.7,0,128C0,57.3,57.3,0,128,0C198.7,0,256,57.3,256,128z};
    \fill[white] svg{M86.3,186.2H70.9V79.1h15.4v48.4V186.2z}
                 svg{M108.9,79.1h41.6c39.6,0,57,28.3,57,53.6c0,27.5-21.5,53.6-56.8,53.6h-41.8V79.1z M124.3,172.4h24.5c34.9,0,42.9-26.5,42.9-39.7c0-21.5-13.7-39.7-43.7-39.7h-23.7V172.4z}
                 svg{M88.7,56.8c0,5.5-4.5,10.1-10.1,10.1c-5.6,0-10.1-4.6-10.1-10.1c0-5.6,4.5-10.1,10.1-10.1C84.2,46.7,88.7,51.3,88.7,56.8z};
  }
}

\newcommand\orcidicon[1]{\href{https://orcid.org/#1}{\mbox{\scalerel*{
\begin{tikzpicture}[yscale=-1,transform shape]
\pic{orcidlogo};
\end{tikzpicture}
}{|}}}}

\newcommand{\verspt}{\\[0.1em]}
\newcommand{\verspb}{\\[0.2em]}
\usepackage{enumitem}
\setlist[itemize]{noitemsep, topsep=0pt}
\usepackage{hyperref} %<--- Load after everything else

\title{Open-VERSO: a vision of experimentation infrastructures beyond 5G, hurdles and challenges}
\author{
Angel Martin \orcidicon{0000-0002-1213-6787}, Researcher at Vicomtech \textit{amartin@vicomtech.org} \and 
Pablo Losada \orcidicon{0000-0002-5593-2716}, Researcher at Gradiant \textit{plosada@gradiant.org} \and 
Carolina Fernández \orcidicon{0000-0003-1865-7177}, Researcher at i2CAT \textit{carolina.fernandez@i2cat.net}
}

\usepackage{xcolor}
\usepackage{svg}

\usepackage{fancyhdr}% http://ctan.org/pkg/fancyhdr
\usepackage{titlesec}
\titleformat{\paragraph}
{\normalfont\normalsize\bfseries}{\theparagraph}{1em}{}
\titlespacing*{\paragraph}
{0pt}{3.25ex plus 1ex minus .2ex}{1.5ex plus .2ex}

\begin{document}
\maketitle
\begin{abstract}
5G led to a digital revolution for networks by leveraging virtualisation techniques to manage software-based network functions through provided standard interfaces, which have matured recently for cloud infrastructure that is widely employed across domains and sectors. This undiscovered potential to adequately respond to concurrent and specialised traffic demands is promising for a wide spectrum of industries. Moreover, it exposes the networking ecosystem to prospects beyond the traditional value chain. However, the configuration, deployment and operation of a 5G network are challenging. Thus, different scientific and research entities have built their own open, evolvable and updateable testbed infrastructure that can be used for experimentation purposes. Such testbeds enable different stakeholders to integrate new systems or features exploiting new technologies, assess the performance of innovative services, and customise operation policies to find optimal setups from a cost-effective perspective. Furthermore, federations of infrastructure allow for wider and more complex experiments to be performed in distributed domains. However, numerous technical and procedural obstacles exist during the building of 5G network testbeds. In addition, some technical barriers persist despite the testing of alternatives and ongoing efforts within open-source systems and commercial equipment portfolios. All these limitations and challenges are relevant for experimenters and stakeholders as they attempt to determine the scope of 5G set expectations.
\end{abstract}

\section{Open-VERSO vision}
5G technologies and networks have gained significant importance in scientific communities, research bodies, and industrial sectors owing to their capacity to push the limits of communications, thus connecting systems and users with unmatched performance capabilities. The use of mobile networks by actors beyond those in the traditional ecosystem, fuelled by virtualisation and softwarisation paradigms, has attracted entities with expertise in cloud infrastructures and data computing technologies that can integrate their solutions into the new, open, and programmable networks.

Testbeds are essential for experimentation and generating prototypes for the burgeoning research that is being applied to controllable 5G technologies and infrastructures. They provide a controlled environment for designing, integrating, deploying, playing, training, and testing new solutions or services. This allows for conclusions to be drawn, solutions to be replicated and optimized, and iterations to be made on a larger scale. Testbeds not only are focused on the testing of innovative applications using an advanced network but also provide a sandbox for network managers to operate a testing infrastructure before migrating solutions to produce or connect with new infrastructures and administrative domains. Furthermore, testbeds enable network engineers to focus on the development of Artificial Intelligence (AI)-based solutions and Machine Learning (ML)-driven techniques, as well as monitored metrics, which automatically adapt the network to the upcoming traffic demands in a smart manner.

Open-VERSO is a Spanish project that has built an open, virtualised technology demonstrator for smart networks. The testbed has been built by three research centres, \href{https://www.vicomtech.org/en}{Vicomtech}, \href{https://www.gradiant.org/en/}{Gradiant} and \href{https://i2cat.net/}{i2CAT}, and employed to accelerate the evolution of 5G and the next-generation mobile communication networks. A federation is of high importance when prototype setups represent traffic crossing multiple networks and systems deployed in distributed infrastructures.

The project participants have developed a set of proof of concepts to validate the deployed infrastructures. These proof of concepts allow for the reproduction of the following different scenarios within the testbeds:
\begin{itemize}
    \item Cells with different base stations and cores in mobility contexts,
    \item AI orchestration solutions for self-organising networking (SON) of radio equipment,
    \item Application programming interfaces (API) to apply Network Slicing configurations, 
    \item Intent-based operations for zero-touch network and resource management, and
    \item Multi-layer monitoring and life-cycle management of distributed systems in multi-tenant edges.
\end{itemize}

However, during the production of the resulting experimentation network and the execution of vertical demonstrators that use the network, certain issues, barriers, and challenges were detected and identified. This could be beneficial for the 5G domain stakeholders when considering building or using a 5G testbed.

This paper is structured as follows. Section \ref{sec:back} lists different aspects to consider when planning to build a new mobile network testbed or to experiment on it. Section \ref{sec:pocs} provides some vertical applications tested in the context of Open-VERSO to exemplify the scope of application of the resulting testbed. Further, Section \ref{sec:6g} provides and overview of the kind of technologies and applications that 6G will target to outline some future requirements for the testbeds. Section \ref{sec:lessons} analytically discusses a set of problems observed during the construction of the testbeds and their federation or the multi-tenant execution of demonstrators. Section \ref{sec:forecast} provides the vision of the consortium on the challenges that will be addressed by software solutions or equipment releases and others that will persist and would need to be addressed with new solutions and advanced systems. Finally, Section \ref{sec:conclusions} provides a few highlights.

\section{Background}
\label{sec:back}

First, this paper lists different aspects to consider when designing, building, deploying, and operating a 5G testbed. These aspects could be considered while designing 6G technologies. To provide a comprehensive review, both technical and normative points are highlighted. Furthermore, this paper provides insights into other European initiatives that stakeholders could use to assess available testbeds. Moreover, these initiatives are relevant to be tracked and studied.

\subsection{Regulation}

When planning to deploy a 5G testbed, one must consider the regulation on spectrum utilisation.

The spectrum and bands available for 5G networks are mainly assigned in auctions held by the national administrations in each country \cite{national5G}. This means that most of the bands allocated for 5G radio networks are exploited through temporal licenses owned by Mobile Network Operators (MNOs) and carriers \cite{spectrum5G}. Although it is possible to reach an agreement with an MNO to locally and temporarily use a slice of their assigned bands; however, such agreements restrict usage of the bands to research and limit their use.

Alternatively, one could apply to the national administration managing the spectrum for a temporal, non-renewable license to use a free band not allocated yet for commercial use in the planned location. With this option, three major concerns should be considered: First, as it is non-renewable, the permission can be revoked at any moment discontinuing its use. Second, as the equipment, antennas, filters, signal amplifiers, remote radio heads (RRHs), base stations and other setup usually support a set of specific bands, a potential move to another band would mean the purchase or acquisition of new equipment that would need to be calibrated. Third, once the band comes up for auction, the experimentation and research licenses are discontinued.

The best alternative is to use a spectrum allocated for non-public networks (NPNs). In this case, the private networks in an area can self-organise to use a set of bands. However, the bands for this purpose change in each country and are not yet decided or allocated in some countries \cite{report5G}.

In any case, the operation of experimental and research networks or NPNs is subject to restrictive regulations in terms of coverage area (power to use in the cells) and the proximity to critical facilities areas such as hospitals and military infrastructures.

Depending on the band available, the ability to cover a wide area or to penetrate indoor infrastructures can vary significantly. Moreover, depending on the free bandwidth, the Key Performance Indicators (KPIs) would enable a powerful mobile network or a poor capacity.

% (PL) In the case of Open-VERSO, the second approach was chosen, and the three research centers requested a segment of the n77 band. As a result, the government granted 40 MHz around 3.975 GHz to each of the three centers, which allowed for different outdoor experiments to be carried out. The results were published in a joint paper [ref].

\subsection{Technologies}

The selected band and bandwidth will guide testbed design as the equipment to use is different.

There are several \textbf{radio} equipment solutions available, including open-source Software-Defined Radio (SDR) options, commercial stacks from telecom industry players, and experimental commercial equipment such as Amarisoft and AWS \cite{amari}. Open-source options include solutions like Open Air Interface (OAI)\cite{oai}, OpenRAN and OpenLTE \cite{openlte}, which run on general-purpose processors and COTS hardware platforms with RF platforms like National Instruments/Ettus Research \cite{ettus} Universal Software Radio Peripheral (USRP), Nuand BladeRF \cite{blade}, and Lime \cite{lime} SDR platforms.

Some of the aspects that need to be considered when choosing from these options are given below:
\begin{itemize}
    \item \textit{Configurability}. Customisation, tuning and settings is capable for open-source solutions; however, options become limited as the commercial setup is selected.
    \item \textit{Updates and costs}. The support of specific features from a release comes with higher pace and early adoption in Open Source and experimental setups, while the commercial setups provide features as they integrate in their massive deployments. For sure the commercial options means higher costs and licenses which need a large deployment to ensure the return of investment. The experimental equipment commercialised are small providers and companies which are more transparent in the roadmaps and already include support in the annual license.
    \item \textit{Stability}. The optimisation of KPIs and configurations is more advanced in commercial than experimentation solutions ready to be installed in outdoor setups and contributes to higher steadiness and performance. Moreover, the availability of experimentation infrastructures is limited, and such infrastructures require periodic reboots and present ineffective and unsupported configurations.
    \item \textit{Interoperability}. Interoperability is a key feature, as commercial solutions tend to be monolithic, which hinders the complete integration with other systems and limits their expansion. However, as they are compliant with external mandatory interfaces, they can be integrated with other equipment. Some experimental systems, however, need to be modified or require adhoc configurations to work with other systems.
    \item \textit{Debug and Monitoring}. Open-source systems already have widely employed tools that can capture, export, and visualise data and logs. Commercial yet experimental equipment include interfaces that export metrics and logs in real time, and can be integrated with cores systems of the testbed, requiring the development and deployment of a broker module. In contrast to this open architecture, commercial stacks favour the onboarding of default systems of the suite to configure, operate, and monitor a solution, limiting their use in a testbed.
\end{itemize}

% PL
Some of the SDR platforms currently supported by open-source RAN frameworks can also be effectively utilized for implementing \textit{hardware offloading} techniques. These techniques involve accelerating computationally heavy functions or blocks in the gNodeB (gNB) physical layer through dedicated hardware co-processors that are typically implemented on Field-Programmable Gate Arrays (FPGAs) or purpose-specific logic devices. Examples of such blocks include the Fast Fourier Transform (FFT) and Inverse Fast Fourier Transform (IFFT) and the Low-Density Parity Check (LDPC) decoder. These blocks are highly parallelizable, allowing co-processors to speed up their execution speed by several orders of magnitude. Tools like the Radio Frequency Network on Chip (RFNoC) framework provided by Ettus simplify the process of offloading these demanding blocks onto the programmable logic of high-end USRPs. However, it is important to note that current open-source frameworks like OAI do not natively support these features. 

The same solutions are applicable to the \textbf{Core} segment of a mobile network. Other alternatives focus just on the Mobile Network Core to provide open-source implementations of Evolved Packet Core (EPC) and 5G Core (5GC), such as Open5Gs \cite{open5gs} and Free5GCore \cite{free5gc}. Other commercial software implementation, not open source and with limited support and updates, is Open5GCore \cite{open5gc}. Furthermore, most of the open source or experimental commercial solutions provide cell/base station and core in a single package or equipment. They can be combined with other instances to produce different cells, several base stations and a single or multiple cores. However, commercial solutions are designed for large deployments and also require a significant investment on connections.

The following aspects should be consider when choosing from these options:
\begin{itemize}
    \item Supported 5G modes. Progress has been made in the development of 5GC modules to enable the transition from the EPC implementation.
    \item Compatibility. Ability to interconnect and operate cells and base stations from other providers.
    \item Scalability. Capacity to manage several base stations and to coexist with another instances with different configurations or ID.
\end{itemize}

The decision in the Radio setup is coupled to the configuration employed in the Core. However, it is important to emphasize that, regarding the \textbf{5G modes} and options, the availability of a 5G \textit{Stand Alone} (SA) implementation is not provided by commercial vendors as their main clients, MNOs, are not providing 5G SA to their subscribers; they are providing 5G \textit{Non-StandAlone} (NSA). Due to the substantial deployment demands and investment involved, MNOs are prioritizing the initial deployment of 5G New Radio (5G-NR) using evolved packet core (EPC) inherited from the long term evolution (LTE) generation. On the other hand, the widespread deployment of 5GCs by MNOs is still limited \cite{report5G}. By contrast, Open Air Interface and Amarisoft have been providing implementations of 5G standalone (SA) setups for several years now.

The \textbf{resources virtualisation framework} is an important technology for realising the 5G stack from European Telecommunications Standards Institute (ETSI) in terms of the Network Function Virtualisation (NFV) infrastructures and its Management and Orchestration (MANO). This virtualisation infrastructure could host the network core(s), when software solutions are employed, the Multi-Access Edge Computing (MEC) platform for third-party services deployment, Business Support Systems (BSS) or Operational Support Systems (OSS), monitoring and visualisation frameworks, data lakes, and any building block supporting the MANO system to be dynamic, intelligent and efficient realising the Self-Organising Networking (SON) paradigm. Thus, the virtualisation infrastructure could also host Artificial Intelligence (AI) or Machine Learning (ML) modules and policy engines to support decision making to apply or modify new configurations at managed logical networks (\textit{network slices}).
% CF: we should not limit ourselves to container-related tools
% Here OpenStack \cite{os}, Kubernetes \cite{k8s} and OpenShift \cite{oshift} are the main container frameworks.
OpenStack \cite{os}, Kubernetes \cite{k8s} and OpenShift \cite{oshift} are some of the most common stacks used to provide virtualisation and orchestration of both containers and Virtual Machines.
In any case, the open source solutions for the Radio and the Core are compatible with the options listed above. Commercial solutions, however, are not, as vendors provide built-in virtualisation stacks that cannot be easily integrated into an adhoc virtualisation infrastructure. In the case of commercial yet experimental solutions, specific license options will be compatible with the chosen virtualisation framework, but only specific versions of Operating Systems (OS) with which products have been validated can be used. Commercial solutions for virtualisation include Amazon EC2, Microsoft Azure or Google Anthos. However, it is preferable for a testbed to be located close to the rest of the physical equipment to ensure minimum latencies and full control to the extent possible.

The following factors should be considered when choosing an option:
\begin{itemize}
    \item Integration. It should be possible to integrate the chosen framework with other management systems.
    \item Isolation. The chosen framework should allow for the isolation of systems for different logical networks and experimentation users.
    \item Interconnection. It should be possible to interconnect the framework with on premise cloud, or external infrastructures to favour experimentation over distributed setups.
\end{itemize}

In terms of the \textbf{Orchestrator}, different options can be considered, such as Open Source Mano (OSM) \cite{osm}, ONAP \cite{onap} or cloudify \cite{cloudify}. OSM is widely employed in 5G Infrastructure Public Private Partnership (5GPPP) projects and has a big community in the network industry, while the other options are transversal and can be applied to other sectors or domains. For these options, any advanced feature that is needed can be developed, trained, and integrated on top of these frameworks. This is a strategical approach taken by commercial vendors who offer pre-configured solutions that lack flexibility for evolution or expansion. Therefore, it is challenging for the testbed to evolve, and the influence of different configurations on the testbed is more difficult to determine.

The following aspects need to be considered when choosing from these options:
\begin{itemize}
    \item Versatile. The chosen framework needs to be integrated with other administrative systems to trigger diverse, global or local actions.
    \item Central. The chosen framework should have the ability to comprehensively manage and monitor the entire network's status.
    \item Interfaced. Interfaces that enable operations from other administration systems should be available.
\end{itemize}

% As explained before virtualisation is part of the network revolution brought by 5G, the transformation of network functions from firmware and equipment into software instances with programmable interfaces enables infinite possibilities for control according to monitored data and management policies.
As stated above, virtualisation and orchestration are key features that further popularise the network transformation required by 5G, which offloads network logic from firmware and equipment into software instances with programmable interfaces enables infinite possibilities for control according to monitored data and management policies.
Software-Defined Networks (\textbf{SDN}s) provide a control plane to program the switches forwarding the packets according to instructions applied on flow tables. Here, the southbound API is employed to interact with the network switches to program the forwarding behaviour. OpenFlow has become the reference for this southbound API \cite{openflow}. To implement the controller of the SDN three main alternatives are available. OpenDayLight (ODL) \cite{odl} could enable the operation of switches using OpenFlow. However, the programming is complex, and the documentation, tutorials, and guides are poor. Moreover, it is difficult to integrate it on working setups, and when issues arise, the debugging and log tools are non-existent. OpenVSwitch (OVS) \cite{ovs}, which can be instantiated on top of OpenStack infrastructure, could be used. However, its capacity for being reconfigured is limited. This means that each OVS can be configured individually for a local configuration, but it is difficult to program several OVS instances at once in a global setup. Furthermore, management services would struggle to apply the obtained results to several OVS, needing local translators for each local setup. Another possibility is to use ONOS \cite{ONOS}. However, it is stacked on top of different description formats and protocols, it may encounter limitations in applying dynamic configurations to switches within a global setup.

Some aspects to consider when choosing from these options are as follows:
\begin{itemize}
    \item Model syncing. The selected framework should be capable of effectively applying the resulting configuration, either on a global or local scale, derived from orchestration. Moreover, it should accommodate a model that comprehensively describes the entire network and its modifications from a global perspective.
    \item Smooth. The selected option should allow for the application of new configurations without the need to reboot or interrupt the forwarding function.    
\end{itemize}

For the \textbf{MEC} the virtualisation platform selected for all the containers to be deployed in the infrastructure is usually employed. To isolate network management, monitoring or virtualised functions from services which could come from third-parties a different namespace and higher constrained security rules are applied.

In terms of User Equipment (\textbf{UE}), the options and possibilities are high. Handheld devices working with high performance and stability are compatible with most of the Radio and Core equipment; additionally, they are appropriate for portable, mobile and reliable setups. They can be plugged with USB into a laptop, PC, or mini-PC running a client application or a service of a server where connectivity is provided using USB tethering reducing overheads, latency or unnecessary bottlenecks when sharing connectivity with WiFi hot spot. It is important that smartphones sometimes require root permissions to use experimentation Subscriber Identity Module (SIMs) or to be subscribed to networks with non-commercial identifiers. Experimental UEs from Quectel, Sierra Wireless, Sunwave or Telit vendors provide evaluation boards with Qualcomm chip sets ready to be plugged via USB to a system without mobile network interface. These devices provides standard configuration interfaces and a considerable amount of logs to achieve a low-level status. However, their delivery times, price, compatibility with specific bands, lack of visibility with regard to firmware updates, lack of optimisation, poorer performance than mobile devices, and their fragility should be considered. To test correct configuration, the easiest setup for experimental commercial setup is to use a twin: one device for the Base Station (BS) and another for the UE. As equipment shipping a BS can be configured to act as a UE, compatibility is ensured. However, this setup is more expensive than using UEs.

Some aspects to consider when choosing from these options are as follows:
\begin{itemize}
    \item Portability. Almost all the possibilities are portable and compatible with experiments on the move. However, mobile devices include batteries for long experimentation sessions and more durable manipulation.
    \item Debugging. When problems arise, the possibility to force configurations and debug from logs can make the difference.
\end{itemize}

\subsection{Initiatives}

% \hl{Experimentation testbeds potentially comprising federated infrastructures \\
% 5GPPP projects \\
% Fed4Fire+ \\
% SNS Streams}

This section provides an overview of the existing mobile network testbeds for experimentation including those developed in European projects, such as 5GPPP or Smart Networks and Services Joint Undertaking (SNS JU), targeting scientific and industrial experimentation. Table \ref{tab:testbeds} show a list of testbeds and their advanced features so that their experimentation potential can be understood.

% CF: provide homogeneised maps (e.g. using https://www.mapchart.net/europe.html)
% since we will probably not have the map from the latest projects
% CF: change the structure of the tables to provide a better way to read the features, e.g. per project:
% first row with title + URL
% second row with centred map
% third row with list of features (try to be homogeneous)

\begin{longtable}[c]{p{\textwidth}}

\hline \verspt

\textbf{Experimentation Testbed Initiative}

\verspb \hline \verspt

\textbf{Open-VERSO} (\url{https://www.openverso.org/en/}) offers an interconnected set of testbeds across Spain, spanning Galicia (GRADIANT), Basque Country (Vicomtech) and Catalonia (i2CAT). It focuses in the "Open RAN" concept and integration on a neutral infrastructure including an advanced network architecture.
\begin{center}
\raisebox{-\totalheight}{
  \includegraphics[width=0.6\linewidth, height=50mm]{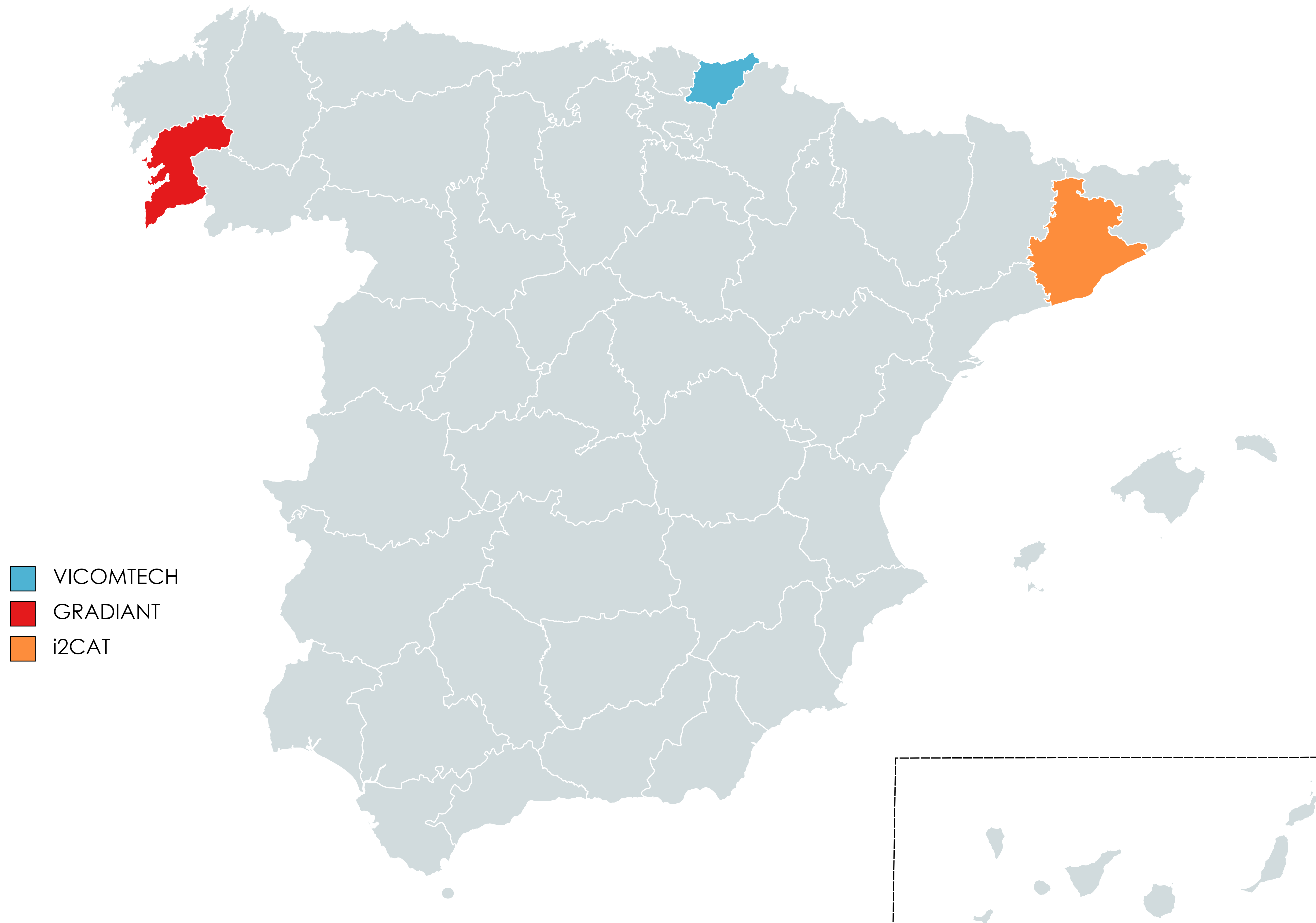}
}
\end{center}\\
\textbf{Features}: E2E including UEs; RINA stack; Spectrum Awareness; In-band full-duplex (IBFD) communications; Load balancing for GPUs at MEC; Multi-RAT sharing; RAN slicing; Inter-cell and inter-public land mobile network (PLMN) mobility; Broadcast technology; Experimentation APIs; Indoor \& outdoor; AI-based MANO; Intent-based actions

\verspb \hline \verspt

\textbf{Fed4FIRE+} (\url{https://www.fed4fire.eu/}) federates interconnected testbeds across several countries in Europe (as well as providing federation with similar initiatives in the USA). It integrates multiple technology domains, such as wired, wireless, IoT, cloud and big data infrastructures, and allows for experiments to be performed on top of federated testbeds.
\begin{center}
\raisebox{-\totalheight}{
  \includegraphics[width=0.6\linewidth, height=50mm]{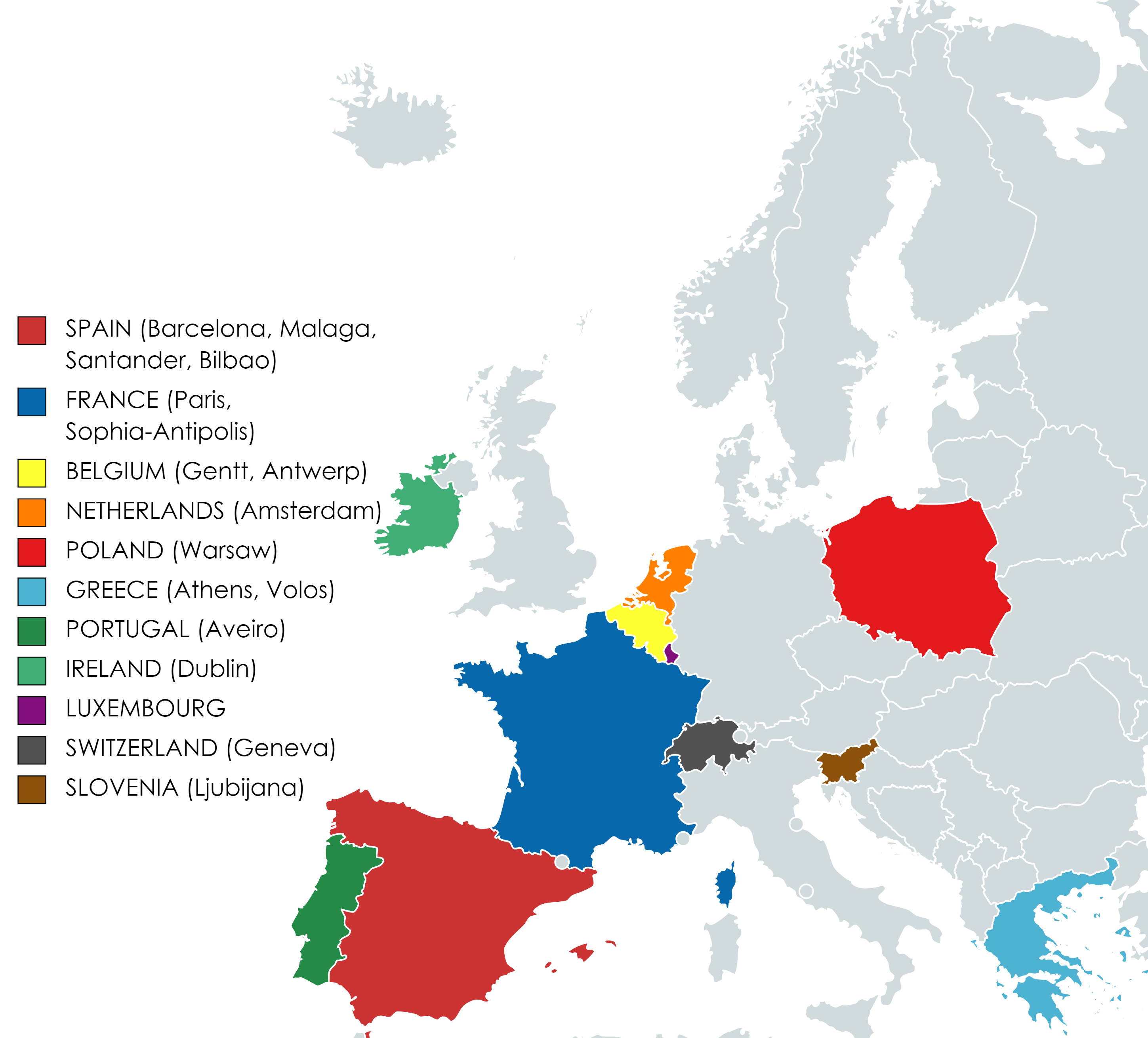}
}
\end{center}\\
\textbf{Features}: Experimentation APIs; Federated experiments; Central broker for experimentation matching; Experiment description language; E2E including UEs; Control on logic domains; Open-calls and hackathons

\verspb \hline \verspt

\textbf{5GBarcelona} (\url{https://5gbarcelona.org/}) is a testbed that connects multiple locations within Barcelona to offer heterogeneous experimentation. Its scope of application is so wide that it is being used for different verticals such as in 5G connected vehicles, particularly in collision situations, smart private networks, emergencies in maritime environments, autonomous robots, augmented reality for online shopping, holographic e-learning, railway IoT systems, and remote surgery.
\begin{center}
\raisebox{-\totalheight}{
  \includegraphics[width=0.6\linewidth, height=50mm]{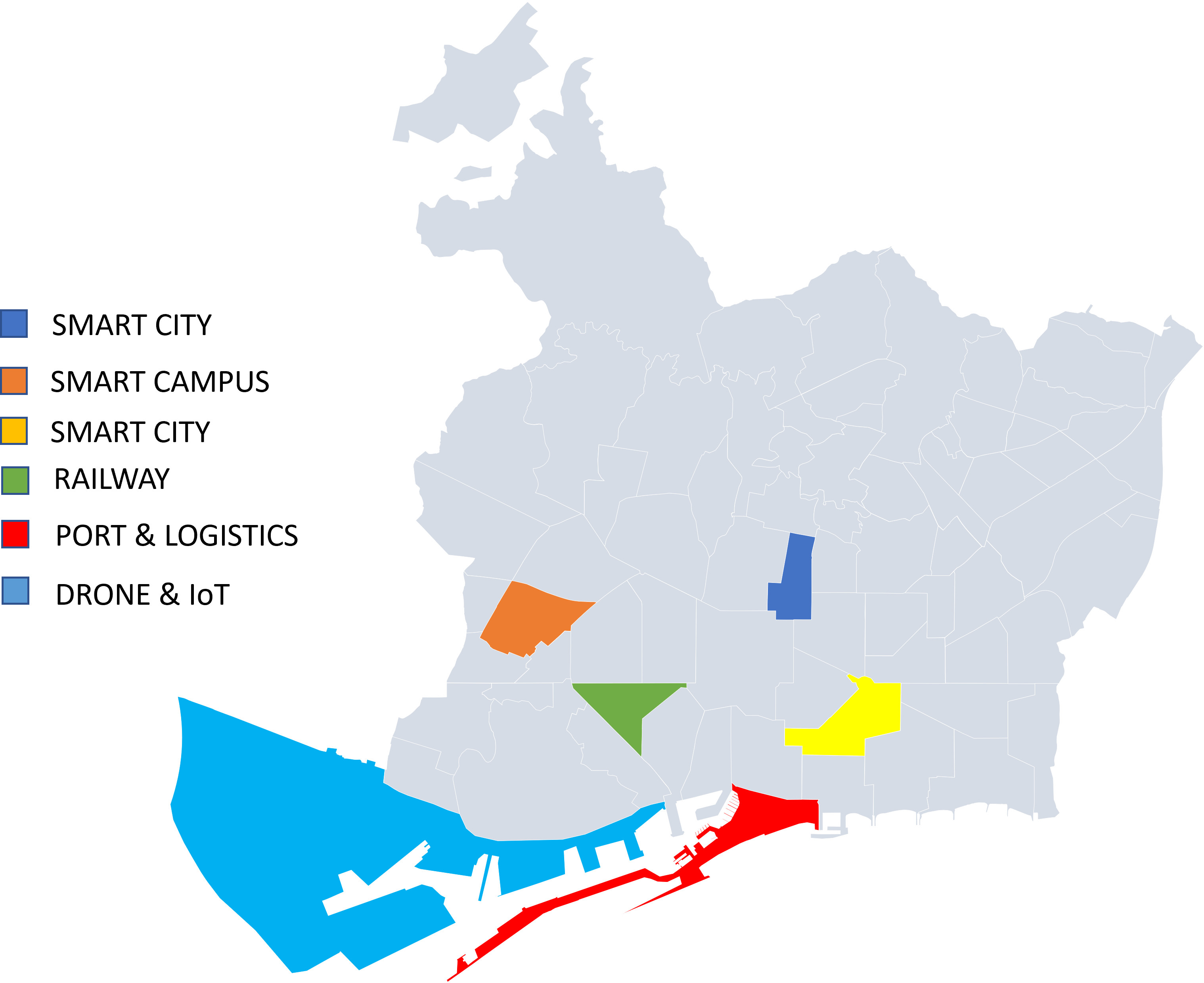}
}
\end{center}\\
\textbf{Features}: Outdoor coverage; Heterogeneous verticals; Hackathons; Open to newcomers

\verspb \hline \verspt

\textbf{5G-EVE} (\url{https://www.5g-eve.eu/}) interconnects multiple testbeds across several European countries. It provides an experimentation testbed for various 5GPPP projects, enabling the testing and implementation of release 16 technologies and key performance indicators (KPIs) across different network slices.
\begin{center}
\raisebox{-\totalheight}{
  \includegraphics[width=0.6\linewidth, height=50mm]{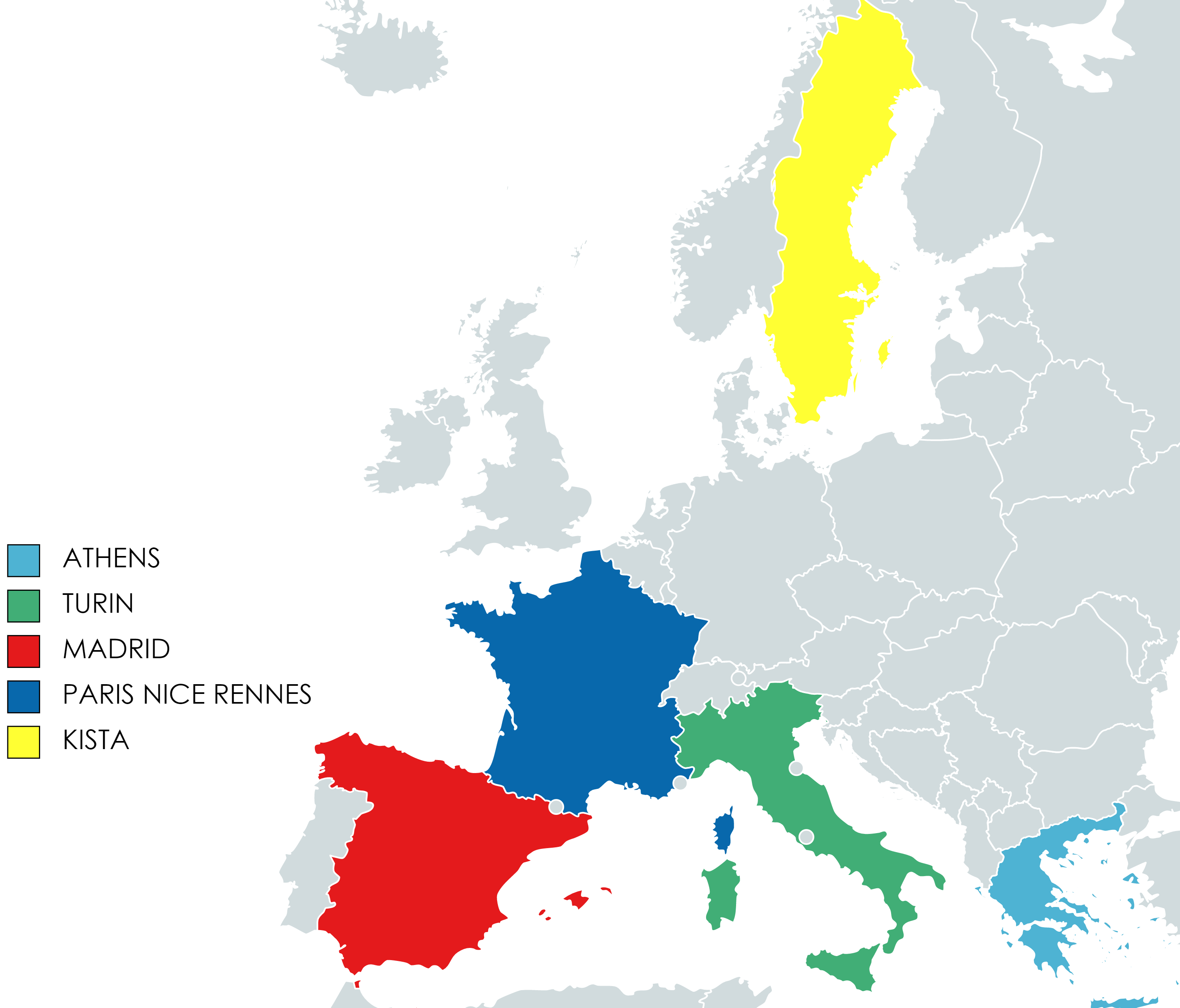}
}
\end{center}\\
\textbf{Features}: Different bands \& spectrum; MEC; Multi-site slicing \& orchestration; Intent-based interfaces; KPI reporting

\verspb \hline \verspt

\textbf{5GENESIS} (\url{https://5genesis.eu/}) interconnects multiple testbeds across European countries. The main applications hosted in the connected testbeds are in the public safety, logistic infrastructures and audiovisual production industries.
\begin{center}
\raisebox{-\totalheight}{
  \includegraphics[width=0.6\linewidth, height=50mm]{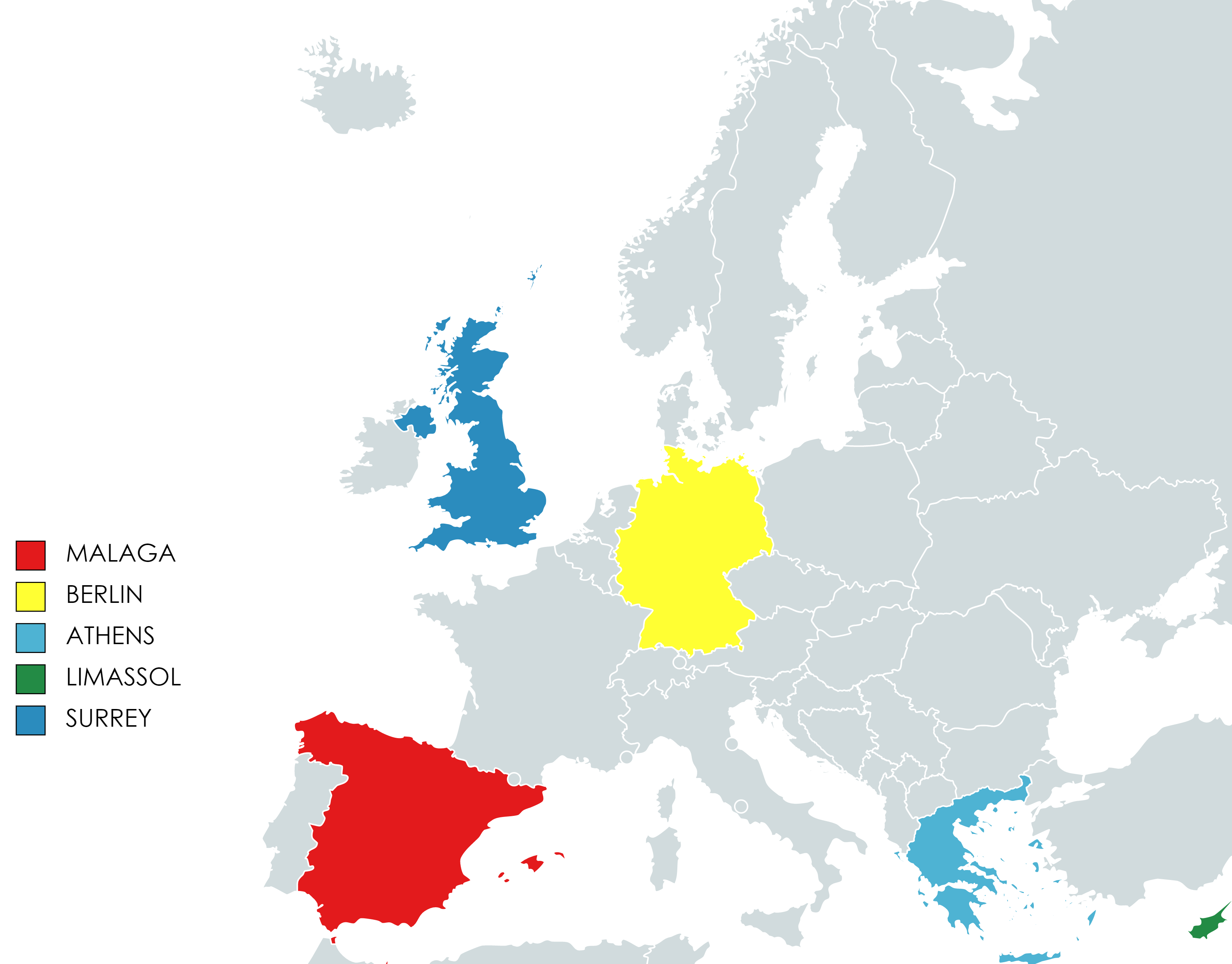}
}
\end{center}\\
\textbf{Features}: KPIs reporting; network slicing; Industrial Internet of Things (IIoT) verticals; MANO automation; satellite communications

\verspb \hline \verspt

\textbf{5G-VINNI} (\url{https://www.5g-vinni.eu/}) interconnects multiple testbeds across several European countries. The resulting infrastructure has been used by 5GPPP projects in the health and first responders domain, such as 5G-HEART, 5GROWTH, 5GSOLUTIONS, and 5G-EPICENTRE.
\begin{center}
\raisebox{-\totalheight}{
  \includegraphics[width=0.6\linewidth, height=50mm]{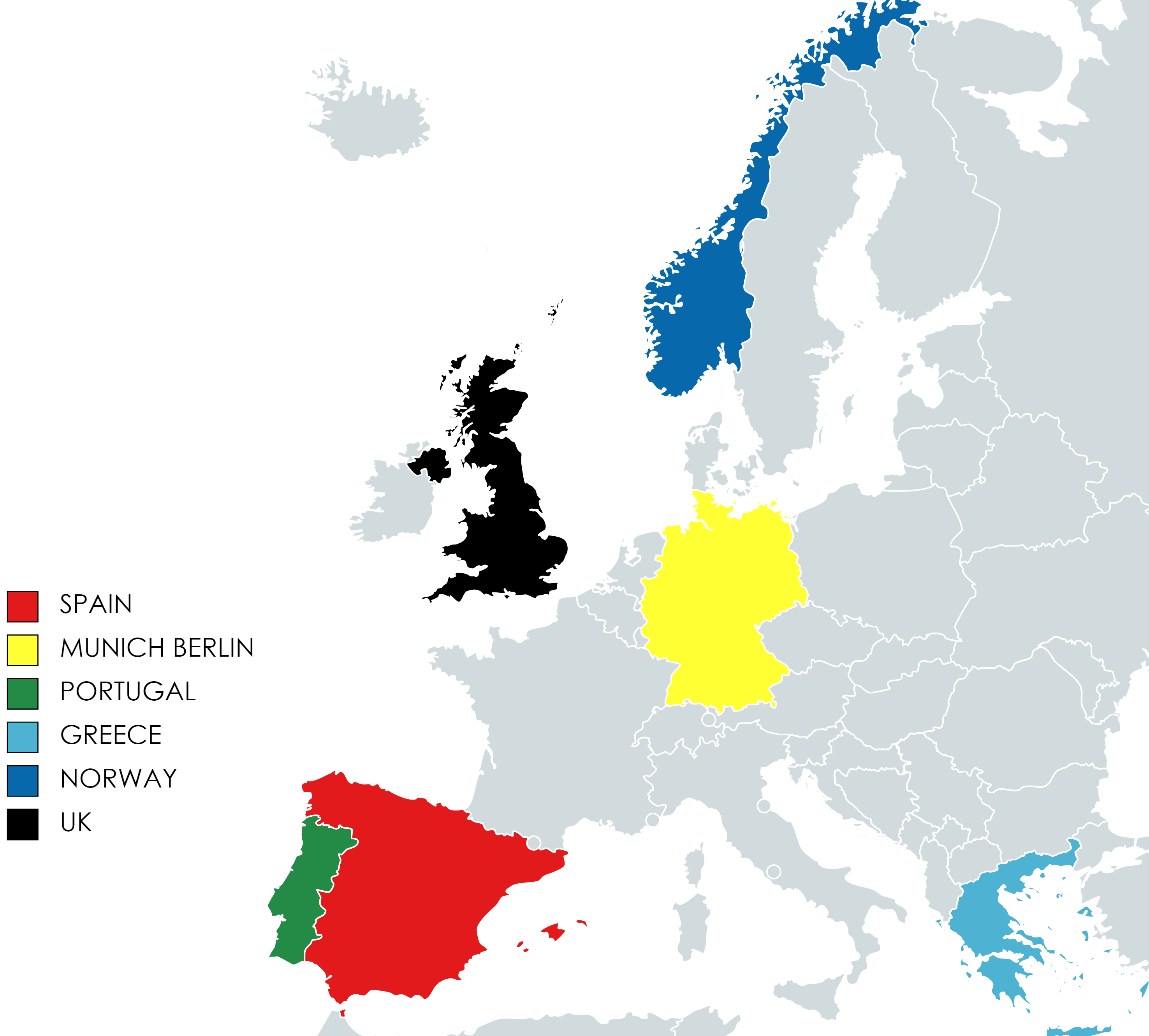}
}
\end{center}\\
\textbf{Features}: E2E including UEs; KPIs reporting; network slicing; experimentation APIs; zero-touch E2E MANO

\verspb \hline \verspt

\textbf{6G-SANDBOX} (\url{https://6g-sandbox.eu/}) interconnects multiple testbeds across several European countries. It targets the "Infrastructure as a Code" paradigm as the Fed4Fire+ project but provides APIs.
\begin{center}
\raisebox{-\totalheight}{
  \includegraphics[width=0.6\linewidth, height=50mm]{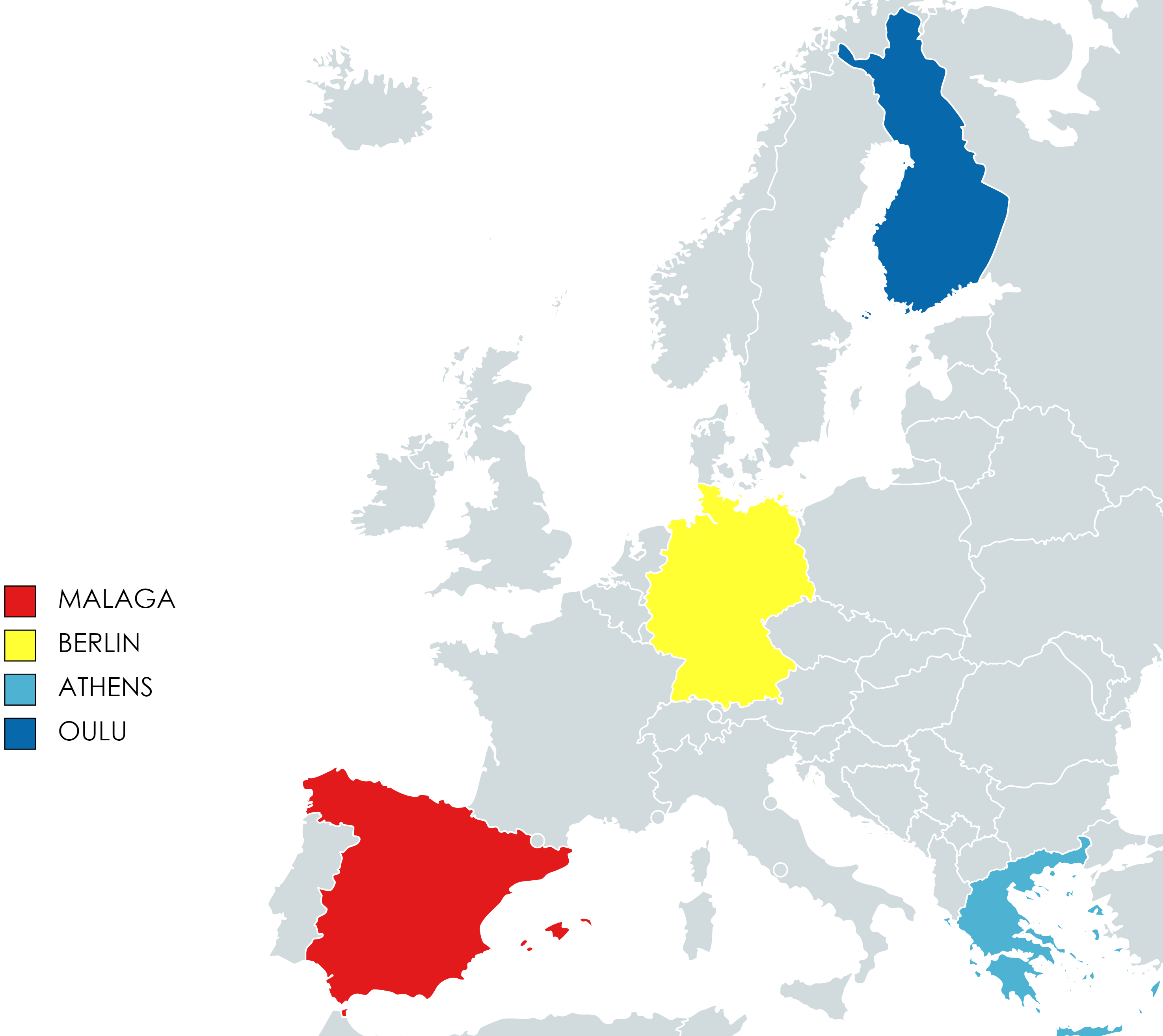}
}
\end{center}\\
\textbf{Features}: E2E including digital and physical nodes; multi-domain resources; gigital twin and XR verticals; open-source components; satellite communications

\verspb \hline \verspt

\textbf{6G-BRICKS} (\url{https://6g-bricks.eu/}) interconnects testbeds between France, Belgium and Greece. It aims to build reusable testbed infrastructures using Open APIs and integrating O-RAN architectures and cell-free and reconfigurable intelligent surface (RIS) scenarios.
\begin{center}
\raisebox{-\totalheight}{
  \includegraphics[width=0.6\linewidth, height=50mm]{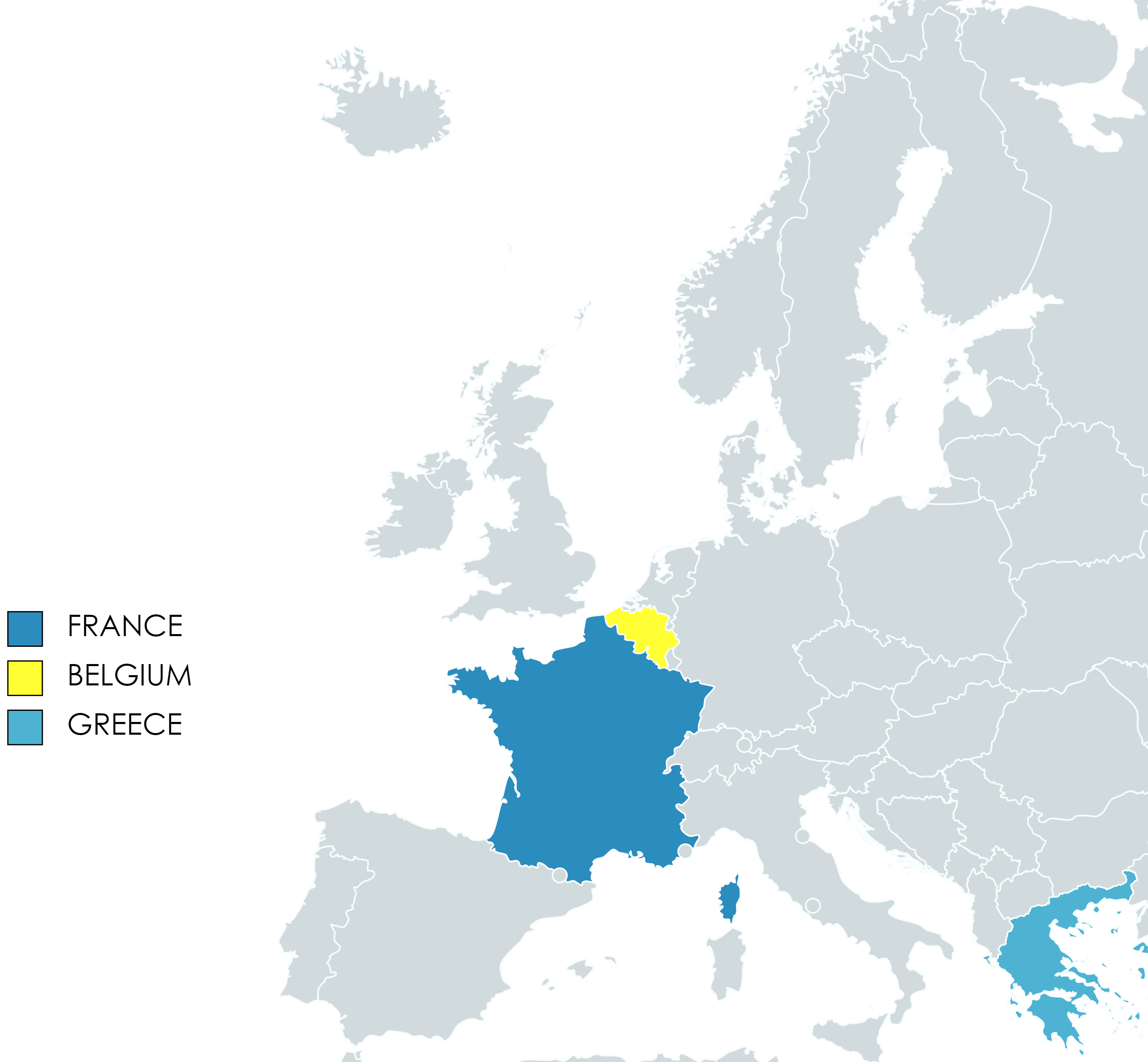}
}
\end{center}\\
\textbf{Features}: Disaggregated control plane and SDN; explainable AI; near-RT radio management and slicing; disaggregated wireless X-Haul; holographic \& digital twin verticals

\verspb \hline \verspt

\textbf{6G-XR} (\url{https://www.6g-xr.eu/}) interconnects testbeds between Spain and Finland. It aims to provide abstraction tools as well as energy measurement frameworks mainly focused on multi access edge computing scenarios.
\begin{center}
\raisebox{-\totalheight}{
  \includegraphics[width=0.6\linewidth, height=50mm]{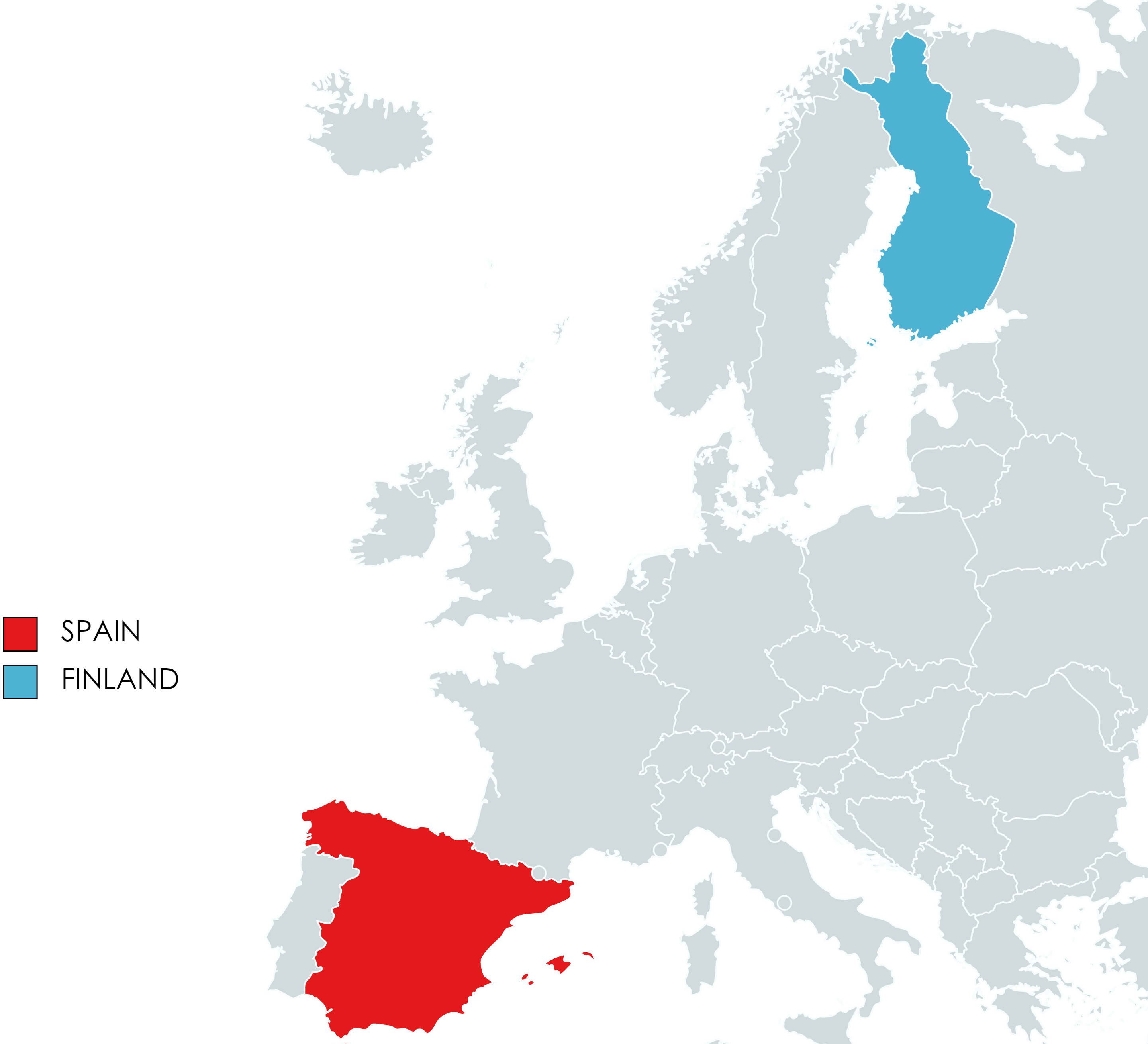}
}
\end{center}\\
\textbf{Features}: Cloud-MEC continuum; spectrum sharing for 5G \& 6G; AI for RAN control \& energy consumption; holographic \& 3D digital twins verticals; E2E energy measurement; open-calls

\verspb \hline \verspt

\caption{\label{tab:testbeds}List of experimentation testbeds of mobile networks, technologies and services.}
\end{longtable}

\section{Deployment scenarios and Use Cases}
\label{sec:pocs}
% \hl{5GPPP Use Cases vs KPIs vs scenarios (eMBB, URLLC, mMTC) \\
% PoCs}

5GPPP and the 5G clusters such as the 5G Automotive Association (AA), Alliance for Connected Industries and Automation (ACIA), European Public Health Alliance (EPHA), European Broadcast Union (EBU) and European Distribution System Operators (EDSO) release reports on the design of representative use cases and studies on the requirements needed to satisfy their communication needs.

The targeted use cases pivot around the main usage scenarios of 5G, Enhanced Mobile Broadband (eMBB), Massive Machine-type Communications (mMTC) and Ultra-reliable and Low Latency Communications (URLLC). Accordingly, the use cases push the required KPIs to values closer to the thresholds outlined by each one or a combination of two of them.

Table \ref{tab:pocs} lists the demonstrators that validate the features included in the Open-VERSO testbed.

\begin{longtable}[c]{p{0.12\textwidth}p{0.2\textwidth}p{0.2\textwidth}p{0.38\textwidth}}

\hline \verspt

\textbf{Name} & \textbf{Description} & \textbf{Experiment goal} &  \textbf{Testbed features}

\verspb \hline \verspt

Broadcast multimedia & Broadcast a video stream from a base station to a UE & Overcome radio capacity with downlink broadcast to SIM-free receivers &
\vspace{-4.5mm}
\begin{itemize}[leftmargin=*]
    \item Virtualisation of SDR systems (BS and UE)
    \item Multi-tenant interconnection of servers and clients in different domains
    \item APIs for remote operation of client containers at UEs and server containers at MECs
\end{itemize}

\verspb \hline \verspt

In-Band Full Duplex (IBFD) communications & Testing of Self-Interference Cancellation (SIC) techniques & Combine and test different SIC algorithms and parameterisations to optimize Self-Interference suppression &
SIC front-end prototype based on a Xilinx RFSoC that includes:
\begin{itemize}[leftmargin=*]
    \item A linear analogue SIC stage
    \item A non-linear digital SIC stage
    \item 40 and 70 MHz QAM signal generator
    \item A set of predefined parameterisations
\end{itemize}

\verspb \hline \verspt

New energy-efficient waveforms for IIoT & Evaluate NOMA (Non-Orthogonal Multiple Access) techniques for massive IIoT communications & Analyse the performance of IIoT links exploiting Sparse Code Multiple Access (SCMA) &
\vspace{-4.5mm}
\begin{itemize}[leftmargin=*]
    \item SCMA communications simulator
    \item Throughput analysis in massive environments
    \item BER and efficiency simulation
    \item Evaluation of channel estimation techniques
\end{itemize}

\verspb \hline \verspt

Social VR & Multi-party experience of telepresence with remote rendering services & KPI-based load balancing of GPU virtualised services in different network Edges &
\vspace{-4.5mm}
\begin{itemize}[leftmargin=*]
    \item GPU virtualisation
    \item Life-cycle manager of VNFs in federated multi-site networks
    \item Automated inventory of computing resources for OS
    \item AI-empowered network topologies
\end{itemize}

\verspb \hline \verspt

Shared Multi-RAT \& sharing & Moving UEs with different KPIs depending on applied policies & Enable APIs to configure RAN quotas and mobility events &
\vspace{-4.5mm}
\begin{itemize}[leftmargin=*]
    \item Intent-based APIs for managing quotas of RAN sharing
    \item Multi-RAT aggregation
    \item Zero-touch network management in multi-tenant environment
    \item Inter-cell and PLMN scenarios
    \item RINA for VNFs over distributed virtualisation infrastructures
\end{itemize}

\verspb \hline \verspt

AI-based Spectrum Awareness & Automatic classification and location of spectral emissions based on Deep Learning (DL) & Detect and locate emissions sources &
\vspace{-4.5mm}
\begin{itemize}[leftmargin=*]
    \item Different DL models for emissions classification
    \item Extensive real-world data sets
    \item Automatic confusion matrix generation
    \item SDR platforms for real signals classification
\end{itemize}

\verspb \hline \verspt

% \caption{\label{tab:pocs}List of experiments deployed and tested on top of Open-VERSO testbed.}
\caption{\label{tab:pocs}Experiments performed on the Open-VERSO testbed.}
\end{longtable}

\section{Enabling technologies for 5G Networks and beyond}
\label{sec:6g}
% \hl{BSS/OSS Discovery \\
% Network Slicing \\
% Cooperative Monitoring Sensing \\
% Zero touch Orchestration \\
% Edge/Cloud continuum \\
% AI-driven SON \\
% Intent Based Actions
% Current Use Cases (5G) vs Future Use Cases (6G) \\
% Hyperconnected \\
% Cross-borders corridors \\
% Radio THz, VLC, Satellital, Underwater \\
% Backhaul Optical \\
% Energy CAPEX/OPEX}

Several technologies have been identified as essential in the current Smart Network and Services (SNS) work programs \cite{sns} outlining the core research directions that will make Europe competitive in the domain of mobile network technologies. Most of these scientific and technological areas have been explored in the 5GPPP phase 1 to 3 calls \cite{5gppp}. Furthermore, they will serve as pillars in next generation mobile networks (6G). Table \ref{tab:6g} provides an overview of the significant technologies encompassing the domains of 5G and 6G experimentation, according to research programs in the European context.

\begin{longtable}[c]{p{0.16\textwidth}p{0.25\textwidth}p{0.13\textwidth}p{0.35\textwidth}}

\hline \verspb

\textbf{Technology} & \textbf{Network Mission} & \textbf{Relevant Publications} & \textbf{Open Challenges}

\verspb \hline \verspt

BSS/OSS Discovery & Automated inventory of new, used and available assets & \cite{ossbss} &
\vspace{-4.5mm}
\begin{itemize}[leftmargin=*]
    \item Automated discovery
    \item Federated infrastructures
    \item Intent-based query
    \item Logical slices management
\end{itemize}

\verspb \hline \verspt

Network Slicing & Create logic instances dedicated and isolated for specific traffic & \cite{slicing1,slicing3} &
\vspace{-4.5mm}
\begin{itemize}[leftmargin=*]
    \item Multi-site orchestration
    \item Multi-domain slice KPIs monitoring
    \item Concurrency policies
\end{itemize}

\verspb \hline \verspt

Cooperative Monitoring Sensing & Capture and correlate data from different levels and players in a network including MANO actions and AI logs & \cite{monitor1,monitor2} &
\vspace{-4.5mm}
\begin{itemize}[leftmargin=*]
    \item Integration with SDR and NFV frameworks
    \item Lightweight probing
    \item Multi-domain push and pull modes for metrics aggregation
    \item Alignment on data namespace and MANO IDs
\end{itemize}

\verspb \hline \verspt

Zero touch Orchestration &  Network automation from the RAN to the core network & \cite{zero1} &
\vspace{-4.5mm}
\begin{itemize}[leftmargin=*]
    \item Universal support for commercial and Open Source setups
\end{itemize}

\verspb \hline \verspt

Edge/Cloud continuum & Migrate services from Cloud to Edge infrastructures to satisfy cost-effective trade-offs & \cite{continuum1,continuum2} &
\vspace{-4.5mm}
\begin{itemize}[leftmargin=*]
    \item Delay-sensitive and context-aware service policies
    \item Session migration
    \item Background instancing for smooth migration when new tenant is ready
    \item Multi-party and multi-network monitoring for multi-path sessions
\end{itemize}

\verspb \hline \verspt

AI-driven SON & ML applied to automated management of cellular networks & \cite{son1} &
\vspace{-4.5mm}
\begin{itemize}[leftmargin=*]
    \item Output models mapped to accepted MANO descriptors
    \item Smooth and intermediate (sub-optimal) transitions to ensure seamless migration
    \item Open and complete data sets for different verticals
\end{itemize}

\verspb \hline \verspt

Intent-based Actions & Translate high-level expectations from users into actionable commands that are transmitted to architectural and networking components & \cite{rfc9315,Clemm2020,9993854} &
\vspace{-4.5mm}
\begin{itemize}[leftmargin=*]
    \item Compromise between generic and extensible models 
    \item Simple VS specific trade-off to foster clarity and granularity
    \item Adoption of NLP technologies
    \item Taxonomies mapping abstract items to heterogeneous components in distributed infrastructures
\end{itemize}

\verspb \hline \verspt

Programmatic overlay networks & High-throughput, programmable L2VPN network overlay on top of the RINA recursive networks & \cite{9815620} &
\vspace{-4.5mm}
\begin{itemize}[leftmargin=*]
    \item Full integration with VIMs (OpenStack and Kubernetes)
    \item N-depth recursion level networks
    \item Support for QoS in every layer
\end{itemize}

\verspb \hline \verspt

Open RAN & Foster interoperability with a variety of multi-vendor ecosystem of open solutions disaggregated from HW & \cite{9124820, 10073507} &
\vspace{-4.5mm}
\begin{itemize}[leftmargin=*]
    \item Agility
    \item Real-time responsiveness
    \item SW APIs for interoperable SW operation
    \item Protect data from management 
\end{itemize}

\verspb \hline \verspt

Shared RAN & Split radio infrastructures and spectrum between multiple operators to reduce costs & \cite{baldesi2022} &
\vspace{-4.5mm}
\begin{itemize}[leftmargin=*]
    \item Coordinate quota policies and economical costs to get cost-effective networks
    \item Monitoring and audit
    \item Dynamic smart contracts
\end{itemize}

\verspb \hline \verspt

Multi-RAT & Combine different radio technologies to appropriately provide connectivity to heterogeneous systems and traffic needs & \cite{s22197591, 9924256, 9557755} &
\vspace{-4.5mm}
\begin{itemize}[leftmargin=*]
    \item Common interfaces to discover, monitor and orchestrate
    \item Traffic classification and prediction to assign more appropriate radio technology to satisfy QoS
    \item Agile and proactive technology migration
    \item Continuity on application and protocols on technology migration
\end{itemize}

\verspb \hline \verspt

Multicast \& Broadcast & Send multicast or broadcast packets from the base station to the subscribed UEs for massive downlink communication & \cite{8971847, 9815676} &
\vspace{-4.5mm}
\begin{itemize}[leftmargin=*]
    \item SIM-free support of networks \& devices
    \item Business model for local or wide hiring of spectrum and network resources to third-parties
    \item Regulation for specific vertical industries
    \item SW solutions for SDR technologies
\end{itemize}

\verspb \hline \verspt

Cross-borders corridors & Mobility, high speeds and long distances will need of advanced and quick cooperation of different networks and domains to provide seamless connectivity & \cite{10025726, 9811139, 10056643} &
\vspace{-4.5mm}
\begin{itemize}[leftmargin=*]
    \item Continuity on application and protocols on mobility
    \item Agile, quick and proactive handover and roaming
    \item Network bonding and multi-SIM concurrent subscription
\end{itemize}

\verspb \hline \verspt

MIMO & Enhance spectral efficiency, capacity and coverage areas with multiple antennas & \cite{7808281, 9318126} &
\vspace{-4.5mm}
\begin{itemize}[leftmargin=*]
    \item Commercial exploitation of the technology
    \item Geographic and timely reconfiguration with radio sensing
\end{itemize}

\verspb \hline \verspt

Full Duplex communications & Increase spectral efficiency and reduce latency by concurrent, in-band transmission and reception  & \cite{Kolodziej2019InBandFT, Wolfe2022ScalableSA} &
\vspace{-4.5mm}
\begin{itemize}[leftmargin=*]
    \item Self-Interference mitigation in high-powered, wideband systems
    \item Self-interference management in phased-arrays
    \item Multi-user interference
\end{itemize}

\verspb \hline \verspt

Non-orthogonal communications & Energy-efficient communications for IoT devices using new physical layer modulations  & \cite{ZilongLiuLiangNOMI, YadavVladislav_Nomi} &
\vspace{-4.5mm}
\begin{itemize}[leftmargin=*]
    \item Optimal modulations to maximize receiver sensitivity
    \item Channel estimation techniques
    \item Multi-user interference mitigation
\end{itemize}

\verspb \hline \verspt

Terahertz Communication & High data rates in line of sight and short distance communications & \cite{8663549, 9473898} &
\vspace{-4.5mm}
\begin{itemize}[leftmargin=*]
    \item Commercial networks operating these bands
    \item Regulation for Non Public Networks
    \item Mechanisms for transition on bands
    \item Use cases with checkpoints for mobility
\end{itemize}

\verspb \hline \verspt

Visible Light Communication & Extremely high capacity integrating space/air/underwater networks & \cite{9208801} &
\vspace{-4.5mm}
\begin{itemize}[leftmargin=*]
    \item Regulation of unlicensed bands
    \item Physical layer technologies
    \item Point to multi-point communication
    \item Full-duplex transmission and reception
\end{itemize}

\verspb \hline \verspt

Satellite Communication &  integration of satellite and terrestrial networks to provide global coverage and seamless connectivity & \cite{9299757, 9616567, 8795462} &
\vspace{-4.5mm}
\begin{itemize}[leftmargin=*]
    \item SW controlled satellite network
    \item APIs for network convergence
    \item Orchestrator for agile provisioning and on-demand resource allocation
    \item Multi-RAT equipment and drivers
    \item Field tests beyond simulations
\end{itemize}

\verspb \hline \verspt

Underwater Communication & Wireless communications adapted to the water medium & \cite{9707771} &
\vspace{-4.5mm}
\begin{itemize}[leftmargin=*]
    \item Specific use cases and operational condition limits
    \item Non-experimental equipment
    \item Field tests beyond simulations
\end{itemize}

\verspb \hline \verspt

Optical Communications for Backhaul & Connection or link of the cell site with the core network needing ultra-low latency and high capacity to avoid bottlenecks & \cite{8650520, 8516977, 9476923} &
\vspace{-4.5mm}
\begin{itemize}[leftmargin=*]
    \item Climate and atmospheric isolation
    \item Wireless/free space optical solutions
    \item Cost-effective alternatives to fiber/fixed/wired setups
    \item Aggregation from different cells
\end{itemize}

\verspb \hline \verspt

Energy CAPEX / OPEX & Inventory of used assets and energy consumption for logical networks (slices) operated for specific applications, users or traffic type and how the provided performance impact on cost and energy & \cite{9576947, 10049023, 7765232, 8894826} &
\vspace{-4.5mm}
\begin{itemize}[leftmargin=*]
    \item Lack of visibility on energy impact when slicing in quotas
    \item Lack of data transparency and audit on impact on network management configurations
    \item Lack of data transparency on energy footprint of traffic 
    \item Presence on energy on all the network management policies
\end{itemize}

\verspb \hline \verspt

\caption{\label{tab:6g}List of relevant mobile network technologies.}
\end{longtable}

It is also important to consider the most representative scenarios considered by 5G projects and industrial clusters and contrast them with the ones dominating 6G space. Table \ref{tab:ucs} provides an overview of the most representative use cases of 5G and 6G reported by working groups and stakeholders \cite{5gpppreport, 5gpppwp}.

\begin{longtable}[c]{p{0.47\textwidth}p{0.47\textwidth}}

\hline \verspb

\textbf{5G Use Case} & \textbf{6G Use Case}

\verspb \hline \verspt

\vspace{-4.5mm}
\begin{itemize}[leftmargin=*]
    \item Automated Guided Vehicles (AGVs)
    \item Real-time monitoring and control of robots in manufacturing
    \item Drones for First Responders
    \item Collaborative Drones
    \item Tele-operation of robots (e.g. remote surgery)
    \item Connected and Automated Mobility (CAM)
    \item Multi-sensor media production and delivery pipelines
    \item Media streaming in Massive Events
    \item Streaming of immersive media formats
\end{itemize}
&
\vspace{-4.5mm}
\begin{itemize}[leftmargin=*]
    \item Multi-party holographic communications and Metaverse-like services
    \item Multi-sensory media experiences (e.g. haptic communications and interactions)
    \item Tactile IoT and Ambient Intelligence
    \item Multi-modal IoT-powered Digital Twins
    \item Connected and Automated Mobility (CAM)
    \item Cobots as robots with symbiotic relations to accomplish complex tasks efficiently
    \item Massive twinning pushing Digital Twin (DT) concept towards a full digital representation
    \item Agriculture IoT in sparse and rural areas
    \item Symmetric distribution in energy grids
    \item Satellite Internet for offshore locations
\end{itemize}

\verspb \hline \verspt

\caption{\label{tab:ucs}List of representative use cases.}
\end{longtable}

From the previously listed enabling technologies and key use cases, Table \ref{tab:ovsota} shows the list of publications developing solutions and technologies on top of the Open-VERSO testbed.

\begin{longtable}[c]{p{0.15\textwidth}p{0.63\textwidth}p{0.14\textwidth}}

\hline \verspb

\textbf{Networking Corner} & \textbf{Title} & \textbf{Publication}

\verspb \verspb \hline \verspt

Architecture & 5G SA Multi-vendor Network Interoperability Assessment & \cite{9547167} \\\
\\ \cline{2-3} \\
& A Cost-Efficient 5G Non-Public Network Architectural Approach: Key Concepts and Enablers, Building Blocks and Potential Use Cases & \cite{s21165578} \\
\\ \cline{2-3} \\
& Cost-Efficient 5G Non-Public Network Roll-Out: The Affordable5G Approach & \cite{9647555} \\
\\ \cline{2-3} \\
& 5G Non-Public Networks: Standardization, Architectures and Challenges & \cite{9611236} \\
\\ \cline{2-3} \\
& A Cloud-Native Platform for 5G Experimentation & \cite{9858299} \\
\\ \cline{2-3} \\
& Prototyping gNBs for Non-Public Networks & \cite{10051296} \\
\\ \cline{2-3} \\
& Application of Multi-Pronged Monitoring and Intent-Based Networking to Verticals in Self-Organising Networks & \cite{9993854} \\
\\ \cline{2-3} \\
& Enabling multi-tenant cellular IoT services over LEO constellations in future 6G networks & - \\
\\ \cline{2-3} \\
& Fundamental Features of the Smart5Grid Platform Towards Realizing 5G Implementation & \cite{10.1007/978-3-031-08341-9_12} \\
\\ \cline{2-3} \\
& Demo: A Decision Support System for Task Offloading Optimization in Cloud-to-Far-Edge Kubernetes Networks & \cite{9915023} \\
\\ \cline{2-3} \\
& Optimal Offloading of Kubernetes Pods in Three-Tier Networks & \cite{9771724} \\

\verspb \hline \verspt

Use Cases & 5G Beyond 3GPP Release 15 for Connected Automated Mobility in Cross-Border Contexts & \cite{s20226622} \\
\\ \cline{2-3} \\
& Adaptive Distributed Beacon Congestion Control with Machine Learning in VANETs & \cite{9751511} \\
\\ \cline{2-3} \\
& Analysis of Vehicular Scenarios and Mitigation of Cell Overload due to Traffic Congestions & \cite{9860570} \\
\\ \cline{2-3} \\
& On Alleviating Cell Overload in Vehicular Scenarios & \cite{10013033} \\
\\ \cline{2-3} \\
& Demonstration and Evaluation of Precise Positioning for Connected and Automated Mobility Services & \cite{9815647} \\
\\ \cline{2-3} \\
& ETSI PoC 5 On-demand Non-Public Networks (NPNs) for industry 4.0: zero-touch provisioning practices in public-private network environments & - \\
\\ \cline{2-3} \\
& 5GCroCo Barcelona Trial Site Results: Orchestration KPIs Measurements and Evaluation & \cite{9815648} \\

\verspb \hline \verspt

vRAN \& ORAN & Realising a vRAN based FeMBMS Management and Orchestration Framework & \cite{9379891} \\
\\ \cline{2-3} \\
& ONIX: Open Radio Network Information eXchange & \cite{9627826} \\
\\ \cline{2-3} \\
& Validating a 5G-Enabled Neutral Host Framework in City-Wide Deployments & \cite{s21238103} \\
\\ \cline{2-3} \\
& VNF Lifecycle Evaluation Study for Virtualized FeMBMS & \cite{9815676} \\
\\ \cline{2-3} \\
& Roadrunner: O-RAN-based Cell Selection in Beyond 5G Networks & \cite{9789832} \\
\\ \cline{2-3} \\
& Nuberu: Reliable RAN Virtualization in Shared Platforms & \cite{10.1145/3447993.3483266} \\

\verspb \hline \verspt

RAT & Software-Defined Vehicular Networking: Opportunities and Challenges & \cite{9281047} \\
\\ \cline{2-3} \\
& Satellite integration into 5G: Accent on testbed implementation and demonstration results for 5G Aero platform backhauling use case & \cite{10.1002/sat.1379} \\
\\ \cline{2-3} \\
& WiMCA: Multi-indicator Client Association in Software- Defined Wi-Fi Networks & \cite{gomez2021} \\
\\ \cline{2-3} \\
& Beam Searching for mmWave Networks with sub-6 GHz WiFi and Inertial Sensors Inputs: an experimental study & \cite{REA2021108344} \\
\\ \cline{2-3} \\
& Delay-Sensitive Wireless Content Delivery: An Interpretable Artificial Intelligence Approach & \cite{9615533} \\
\\ \cline{2-3} \\
& The Satellite Component of VDES: A PHY Layer Implementation Performance & - \\
\\ \cline{2-3} \\
& Air-to-Ground Channel Characterization for Unmanned Aerial Vehicles Based on Field Measurements in 5G at 3.5 GHz & \cite{9941634} \\
\\ \cline{2-3} \\
& Aplicaciones de las técnicas de transmisión In-Band Full-Duplex en el ámbito militar & - \\
\\ \cline{2-3} \\
& 5G-CLARITY: 5G-Advanced Private Networks Integrating 5GNR, Wi-Fi and LiFi & \cite{9722800} \\

\verspb \hline \verspt

Network Slicing & On 5G network slice modelling: Service-, resource-, or deployment-driven? & \cite{PAPAGEORGIOU2020232} \\
\\ \cline{2-3} \\
& Resource Allocation for Network Slicing in Mobile Networks & \cite{9272281} \\
\\ \cline{2-3} \\
& G-ADRR: Network-wide slicing of Wi-Fi networks with variable loads in space and time & \cite{9380999} \\
\\ \cline{2-3} \\
& Blockchain-Based Zero Touch Service Assurance in Cross-Domain Network Slicing & \cite{9482602} \\
\\ \cline{2-3} \\
& Solutions for Traffic Isolation in 5G Infrastructures Using Network Slicing Techniques & \cite{9998422} \\
\\ \cline{2-3} \\
& IWCMC Design of AI-based Resource Forecasting Methods for Network Slicing & \cite{9824551} \\

\verspb \hline \verspt

Orchestration & Time-Sensitive Mobile User Association and SFC Placement in MEC-Enabled 5G Networks & \cite{9427232} \\
\\ \cline{2-3} \\
& Ensuring Session Continuity for Railways Using a Stateful Programmable Data Plane & - \\
\\ \cline{2-3} \\
& vL2-WIM: Flexible virtual layer 2 connectivity services in distributed 5G MANO domains & \cite{9482422} \\
\\ \cline{2-3} \\
& Enhanced Access Traffic Steering Splitting Switching with Utility-Based Decisioning & - \\
\\ \cline{2-3} \\
& Towards Zero Touch Management: A Survey of Network Automation Solutions for 5G/6G Networks & \cite{9913206} \\
\\ \cline{2-3} \\
& OROS: Orchestrating ROS-driven Collaborative Connected Robots in Mission-Critical Operations & \cite{9842799} \\

\verspb \hline \verspt

RINA & ARCFIRE: Experimentation with the Recursive InterNetwork Architecture & \cite{computers9030059} \\
\\ \cline{2-3} \\
& A P4-Enabled RINA Interior Router for Software-Defined Data Centers & \cite{computers9030070} \\
\\ \cline{2-3} \\
& RINA-Based Virtual Networking Solution for Distributed VNFs: Prototype and Benchmarking & \cite{9815620} \\

\verspb \hline \verspt

Monitoring Sensing & Adaptive Rate Control for Live streaming using SRT protocol & \cite{9379708} \\
\\ \cline{2-3} \\
& Deep Learning based Classification of CP-OFDM RAT-Dependent Signals & \cite{9689434} \\
\\ \cline{2-3} \\
& ARENA: A Data-driven Radio Access Networks Analysis of Football Events & \cite{9234430} \\
\\ \cline{2-3} \\
& An ns-3 implementation of a battery-less node for energy-harvesting Internet of Things & \cite{10.1145/3460797.3460805} \\
\\ \cline{2-3} \\
& Multi-access Edge Computing video analytics of ITU-T P.1203 Quality of Experience for streaming monitoring in dense client cells & \cite{10.1007/s11042-022-12537-4} \\
\\ \cline{2-3} \\
& Traffic Classification for Network Slicing in Mobile Networks & \cite{electronics11071097} \\
\\ \cline{2-3} \\
& Adaptive QoS of WebRTC for Vehicular Media Communications & \cite{9828782} \\
\\ \cline{2-3} \\
& Reduced precision discretization based on information theory & \cite{ARES2022887} \\
\\ \cline{2-3} \\
& Deep Learning based CP-OFDM Signal Classification with Data Augmentation & \cite{9858310} \\
\\ \cline{2-3} \\
& Second Order Statistics of N-Fisher-Snedecor F Distribution and Their Application to Burst Error Rate Analysis of Multi-Hop Communications & \cite{9964041} \\

\verspb \hline \verspt

AI for SON & Tiki-Taka: Attacking and Defending Deep Learning-based Intrusion Detection Systems
 & \cite{10.1145/3411495.3421359} \\
\\ \cline{2-3} \\
& Predictive CDN Selection for Video Delivery Based on LSTM Network Performance Forecasts and Cost-Effective Trade-Offs & \cite{9247478} \\
\\ \cline{2-3} \\
& AI-Based Autonomous Control, Management, and Orchestration in 5G: From Standards to Algorithms & \cite{9277894} \\
\\ \cline{2-3} \\
& CPSoSaware\: Cross-Layer Cognitive Optimization Tools \& Methods for the Lifecycle Support of Dependable CPSoS & \cite{9155036} \\
\\ \cline{2-3} \\
& AI-Empowered Software-Defined WLANs & \cite{9422336} \\
\\ \cline{2-3} \\
& AI-Based Autonomous Control, Management, and Orchestration in 5G: From Standards to Algorithms & \cite{9277894} \\
\\ \cline{2-3} \\
& Unsupervised Clustering for 5G Network Planning Assisted by Real Data & \cite{9751769} \\
\\ \cline{2-3} \\
& A PPO Reinforcement Learning MAC Scheduler & \cite{ABBASI2021104234} \\
\\ \cline{2-3} \\
& Predictive Path Routing Algorithm for low-latency traffic in NFV-based experimental testbed & - \\

\verspb \hline \verspt

MEC & 5G MEC-enabled vehicle discovery service for streaming-based CAM applications
 & \cite{Velez2022} \\
\\ \cline{2-3} \\
& Latency and Mobility-Aware Service Function Chain Placement in 5G Networks & \cite{9210756} \\
\\ \cline{2-3} \\
& AI@EDGE: A Secure and Reusable Artificial Intelligence Platform for Edge Computing & \cite{9482440} \\
\\ \cline{2-3} \\
& Assessment of the effects of 5G MEC cache on DASH adaptation algorithms & \cite{9828651} \\

\verspb \hline \verspt

Multi-party and Federation & Multi-Party Collaboration in 5G Networks via DLT-Enabled Marketplaces: A Pragmatic Approach & \cite{9482487} \\

\verspb \hline \verspt

Energy & Bayesian Online Learning for Energy-Aware Resource Orchestration in Virtualized RANs & \cite{9488845} \\
\\ \cline{2-3} \\
& Optimal energy-aware task scheduling for batteryless IoT devices & \cite{9446584} \\

\verspb \hline \verspt

Security & Addressing Cybersecurity in the Next Generation Mobility Ecosystem with CARAMEL
 & \cite{ARGYROPOULOS2021307} \\

\verspb \hline \verspt

\caption{\label{tab:ovsota}List of publications from the consortium on mobile network technologies employed in Open-VERSO project.}
\end{longtable}

\section{Lessons Learned}
\label{sec:lessons}
% \hl{From the point of view of: \\
% Network Operator \\
% Network Federator \\
% Network Experimenter Application or Service \\
% Network Experimenter Network Technology}

% This section provides a summary of all the issues raised during the designing, selecting, deploying, configuring, interconnecting, operating and orchestrating the federated testbeds and while using it for experimentation over the different testbeds.
This section aims to provide a summary of the issues that emerged during the design, selection, deployment, configuration, interconnection, operation, orchestration of federated testbeds, and the experimentation that was conducted over them.

In the Open-VERSO project, the consortium has played the following different roles: network operator of each individual testbed, network federator connecting each testbed to others with pairs, network experimenter integrating and testing network technologies and sharing the result with others to port the same approach to other testbeds, and network experimenter of innovative applications or services, where APIs are mainly used and improved as extra features were identified.

% Table \ref{tab:sw} and \ref{tab:hw} provide an overview of the issues and arranges these per category, underlining the impact and proposing workarounds to be considered and implemented.
Section \ref{section:software} and Table \ref{tab:hw} provide an overview of the emergent issues and categorises them, underlining their impact and proposing workarounds that could be considered and implemented.

% CF: consider adding "Requirements", "Networking"

\subsection{Requirements}

Ideally, all the requirements for software, systems, and use cases should be available and sufficiently mature as soon as possible. As requirement elicitation is an iterative process that typically overlaps with the beginning stages of project deployment and integration, it would be advisable to address any changes in requirements before reaching two-thirds of the project's lifespan, and it is essential to avoid exceeding four-fifths of the project's duration. The former leaves room for developers and infrastructure operators to implement changes that require "medium" and "major" workloads, while accommodating other deviations during the last third of the project's life. If changes in requirements are provided during the last fifth of the project’s life, a no-return period should be entered for such changes, and they should only be considered for "minor" and few "medium" workloads.

In any case, the maximum time within which the final requirements should be provided is determined by several factors, with most of them related to the complexity of the implementation (or arrangement) of the requirements and dependencies on external providers (e.g. third parties providing connectivity, equipment, spectrum licences or consultancy/support in third-party software). A good approach is to characterise the time required to meet the requirements well in advance and naturally plan to prioritise tasks of lower expertise, bureaucratic procedures, or tasks of higher (or unknown) complexity. This can be the case for national requests to acquire radio spectrum licences, tunnels spanning regions, or even National Research and Education Network (NRENs), or tasks whose complexity is yet to be determined, which depend on the (unknown) availability of third-party developers or consultants.

Finally, best practices dictate that some considerations should be taken early during the design phase. These apply to (i) security, (ii) scalability, (iii)  stability, and (iv) minimal throughput and are mostly transversal to the nature of the project, they are typically applicable across different project types, irrespective of whether they involve software components, DevOps deployments, system configurations, and so on. A compromise between the lifetime of the project and the foreseeable capabilities to be offered and their maturity will determine to what considerations should be taken. As stated above, it is crucial to allocate additional buffer time in the planning phase to accommodate changes or extensions.

This means that for sustainable infrastructure valid for a project or for experimenting on top of it, the solutions should be updateable, open, and maintained by a community for open-source technologies, or by a vendor for HW apparel, avoiding situations where the project relies on frozen versions of operating systems (OS) or becomes limited to the local infrastructure in terms of operations.

\label{section:software}
\subsection{Software}

The following subsections analytically discuss the functions performed by the software components and the problems that arose, including their simplifications and workarounds, and alternative software components that were evaluated and integrated into the Open-VERSO testbed.

\subsubsection{Radio management}

\paragraph{Open Air Interface}
To manage an Open Air Interface (OAI), the operator must have some knowledge of UEs, eNBs, gNBs, EPC and 5G cores, and specifically their behaviour in physical layer configurations.

Some of the issues that arose during operation, as well as the approaches or workarounds, are enumerated below>

\begin{enumerate}
\item Poor stability of the software, with unexpected and unpredictable stops during execution and total failure. For long tests to be conducted, the tests can be performed with different Universal Software Radio Peripherals (USRPs) and different configuration values.
\item Owing to the poor throughput of the software, several tests are needed with the configuration files to obtain the expected KPIs for 5G communication, as no documentation has been released on working configurations that realize massive downlink throughput. Additionally, finding a configuration that achieves symmetric performance with an unbalanced setup with prioritised uplink is difficult, as for most of the employed configurations, download traffic takes precedence. Trial-and-error testing is not easy as the instability does not ease the execution of long tests.
\item Conducting tests in new USRPs is challenging as most samples in the community are for LTE technology and older USRP models. Moreover, these models receive the most support from the community.
\end{enumerate}

\paragraph{FlexRAN and FlexRIC}
Both open-source projects promoted by Mosaic5G attemp to leverage the ability of SDR to publish a programmable management of the control and data planes. These projects, such as FlexRIC, which is an evolution of FlexRAN, have become pillars for ORAN owing to their inter-operable software solutions and stacks. Thus, they enable dynamic spectrum sharing scenarios and advanced management of base station configurations to coordinate several cells. Their usage requires deep knowledge of RAN specifications. However, their community is quite active and quickly responds to issues. FlexRIC is also being integrated with different controllers. Moreover, the operation of multiple gNBs tenants concurrently is also a key feature being supported to unleash the full potential of this solution. As these technologies are currently in an early adoption stage, their documentation is not complete, which has limited their application to several instances.

\subsubsection{Infrastructure management}

\paragraph{OpenStack}

When operating OpenStack, the following skills were validated: management of volumes and flavours, configuration of plug-ins and registry, and usage of Heat orchestration templates.

During such operation, several issues had to be confronted. Some examples are enumerated below.

\textbf{Fine-grained deployment}
\begin{enumerate}
\item There was no straightforward way of guiding the placement of virtual nodes based on HW capabilities. The solution was to bind the server characteristics to specific metadata in steps.
% \item A never-ending deployment \& wrong placement were faced when attempting to match to specific compute nodes (i.e. when deploying the VR-related applications). The logs of the OpenStack services did not provide enough information, and in the end the issue was due to the lack of disk in the allocated partition for the OpenStack volumes, which are attached to instances in part of the Open-VERSO setup or the unavailability of the physical GPU when another virtual instance is using it or when it does not exist. Therefore, freeing some space, as well as extending disks and moving partitions in the server with GPU capabilities.
\item A never-ending deployment and wrong placement were faced when attempting to match to specific compute nodes (i.e. when deploying the VR-related applications). The logs of the OpenStack services did not provide enough information because of a lack of disk space in the allocated partition for the OpenStack volumes, which are attached to instances in part of the Open-VERSO setup. Therefore, freeing some space, as well as extending disks and moving partitions in the server with GPU capabilities fixed the issue.
\item Failure during deployment occurred because a physical GPU was in use and therefore unavailable or when it did not exist in the expected server. This is a limitation of the virtualisation of some GPU models, and as such, there is no specific solution other than to track and keep one instance at most using this resource.
\item A bad configuration can make it difficult for incoming traffic to reach inside the infrastructure and nodes deployed with OpenStack. It is important to start with a configuration which already provides incoming and outgoing traffic from/to the Internet.
\item To update the version of running infrastructure, it is usually necessary to tear down the old infrastructure. This is a major milestone in the life cycle of virtualisation infrastructure and should be programmed in advance.
\end{enumerate}

\textbf{Disk-intensive usage}
\begin{itemize}
\item It is not uncommon to run out of space, especially when using VR-related applications with a high disk consumption (e.g. approximately 100 GB per instance) are run. To solve this issue, new partitions were created and volumes were migrated (offline, not through hot migration) to servers with extended disks.
\end{itemize}

\textbf{Frequent configuration adjustments}
\begin{itemize}
\item As changes took place in some network configurations, the endpoints used to expose all registered services needed to be modified. This was achieved by directly updating OpenStack. There was no need to redeploy services.
\end{itemize}

\textbf{Virtual Routing}
\begin{itemize}
\item Employ virtual routers as programmable elements to build a dynamic and mutable topology, as a Software-defined Network is not as agile and dynamic as needed by topology optimisers and routing solutions. This often requires retiring routers and substituting them with new ones. Moreover, the impact of the change is not evident in many cases, and virtual routers need to run for some time before it can be applied to traffic.
\end{itemize}

\paragraph{Kubernetes}

In Kubernetes, most of the issues that arose were related to maintenance activities. Some examples are stated
below.
\begin{enumerate}
    \item To ensure isolation of concurrent experimenters or systems from slices, namespaces should be used when creating clusters and pods. This means that the design needs to consider that from the very beginning to avoid visibility on other slice or experimenter systems.
    \item When rebooting a cluster for maintenance or scaling purposes, the process is not as straightforward as expected. Rebooting takes a substantial amount of time and sometimes error bootstrapping pods unexpectedly occur, which are fixed by restarting the procedure.
    \item Kubernetes was not designed to deal with networking functions. For example, several Container Network Interface (CNI) plugins are not compatible with the SCTP protocol, a key piece of the 5G control plane. The election of the CNI plugin and other complements (such a multus \cite{MultusCitation}) will impact the Container Network Function (CNF) platform operation, so they must be carefully chosen.
\end{enumerate}

\subsubsection{Network Orchestration}

Open Source MANO (OSM) is the NFV orchestrator in use for the Open-VERSO project. To properly configure and maintain it, one needs to be able to perform troubleshooting, development modification and redeployment; moreover, knowledge of the workflow and code base is needed.

During such operation, several issues arose and had to be addressed. Some examples are stated below.

\begin{enumerate}
\item The stability for OSM, which can lead to lack of availability, depends considerably on the version used and the nature of deployment. To overcome such chances, snapshots were taken within OpenStack (as the platform virtualising the OSM and other tools in the management/control plane). Furthermore, scripts for Infrastructure-as-a-Code (IaC) were prepared for easier redeployment and configuration in cases where snapshots were sufficient.
\item The descriptors that determine the logic of the OSM packages have been updated several times across the OSM releases, and sometimes this implies a lack of backwards compatibility. To overcome this challenge, public material was analysed throughout their different stages of development, in addition to directly reviewing issues in the official mailing lists and interactive channels.
\item The utilisation of some baseline systems in the OSM framework such as the Kafka messaging platform and the telemetry (Grafana or ELK) system for any network management purpose beyond the components of the OSM suite is intricate. It is evident that in order to achieve smart and cognitive network management and orchestration, the monitoring and communication of multiple systems based on a shared status or triggers play a pivotal role. However, usually it is necessary to deploy a new IoT messaging platform, a monitoring/telemetry database and a data visualiser to make them accessible by network management systems outside the OSM box.
\item Finally, OSM shows a default behaviour that prevents Intel's Enhanced Platform Awareness (EPA) features, which the OSM's LCM module understands as metadata in the OpenStack flavours during the life-cycle management of the Network Services (NSs), from working. However, this is also the method used in the project to bind given capabilities of clusters (e.g. GPU-capable servers) to the deployment of specific NSs, and because of this, the default behaviour was preventing our logic for a more fine-grained deployment. In the end, this was solved by modifying specific parts of the code, which was possible after reaching out to the DevOps' module owner for detailed explanation and complementing the process with the pipeline files in the repositories to finally redeploy the specific affected module (i.e. LCM).
\end{enumerate}

\subsubsection{Software Defined Networking}
\paragraph{OpenDayLight}
Opendaylight (ODL) was planned to be used by Self-Organising network solutions to apply new configurations such as creating network topologies and update them seamlessly according to identifies or forecasted traffic patterns and demands. Thus, this SDN approach can be used to deploy a cost-effective setup that minimises latencies or avoid bottlenecks. However, the deployment with \textit{mininet}, as a sandbox for that purposes, is feasible but it is not operable at the same level when it is connected to a virtualised infrastructure using OpenStack or OSM. The integration here is not documented and does not produce any result.

\subsection{Equipment}
% \hl{Maturity \\
% Late adopters of 3GPP releases \\
% Timing for delivery \\
% Fragile manipulation \\
% Not optimised when compared to Massive Consumer Devices \\
% Licensing and lack of visibility on roadmap \\
% Firmware updating \\
% Difficulties to operate from drivers \\
% Just Local Monitoring}

Table \ref{tab:hw} shows, the functions performed by the equipment and the problems that arose, including their simplifications, workarounds, and alternatives that were evaluated and integrated to include the hardware component into the Open-VERSO testbed.

\begin{longtable}[c]{p{0.12\textwidth}p{0.23\textwidth}p{0.33\textwidth}p{0.22\textwidth}}

\hline \verspb

\textbf{Equipment} & \textbf{Skills validated} & \textbf{Issues arisen} & \textbf{Workaround}

\verspb \hline \verspt

Amarisoft Callbox Pro &
\vspace{-4.5mm}
\begin{itemize}[leftmargin=*]
    \item Run as UE, eNB, gNB, EPC, 5GC
    \item RAN slicing
    \item Exportable monitoring
    \item Telematic operation
\end{itemize} &
\vspace{-4.5mm}
\begin{itemize}[leftmargin=*]
    \item Virtualisation not possible
    \item Limits on total BW
    \item Limited power and indoor coverage
    \item Interference between TX and RX channels
    \item Lack of visibility on roadmap
    \item Just embedding local monitoring
\end{itemize} &
\vspace{-4.5mm}
\begin{itemize}[leftmargin=*]
    \item Add filter to isolate TX and RX channels
\end{itemize}

\verspb \hline \verspt

AWS RRH &
\vspace{-4.5mm}
\begin{itemize}[leftmargin=*]
    \item Different bands and frequencies
    \item Long times up \& running
    \item Telematic operation
\end{itemize} &
\vspace{-4.5mm}
\begin{itemize}[leftmargin=*]
    \item SW updates on BS will require an SW update in RRH equipment
    \item SW updates require a specific OS version
    \item High costs and long delivery times
\end{itemize} &
\vspace{-4.5mm}
\begin{itemize}[leftmargin=*]
    \item Design and plug a low power antenna, filter and amplifier setup
\end{itemize}

\verspb \hline \verspt

Experimental Modems &
\vspace{-4.5mm}
\begin{itemize}[leftmargin=*]
    \item Compatible with commercial, NPNs, NSA and SA setups
    \item Configurable with AT commands
    \item Usable from different OS
\end{itemize} &
\vspace{-4.5mm}
\begin{itemize}[leftmargin=*]
    \item Fragile evaluation board
    \item Fragile SIM dock
    \item Driver support in specific OS versions
    \item Bands support
    \item SIM security version support
    \item Suboptimal KPIs compared to those of smartphones
    \item Late adopters of 3GPP releases
    \item Firmware updates to the recent 3GPP release is not incremental; thus, some previous features are lost
    \item Just embedding local monitoring and logs
\end{itemize} &
\vspace{-4.5mm}
\begin{itemize}[leftmargin=*]
    \item Have smartphones as a backup to avoid stopping testing sessions
\end{itemize}

\verspb \hline \verspt

Antennas &
\vspace{-4.5mm}
\begin{itemize}[leftmargin=*]
    \item Working in the bands specified to have regular radiation pattern
    \item Easy to purchase even with specific requirements
    \item Weather/Outdoor proof/resist
    \item Anchoring tools available
\end{itemize} &
\vspace{-4.5mm}
\begin{itemize}[leftmargin=*]
    \item Designed for radiation upwards 
    \item Low protection against outdoor conditions
\end{itemize} &
\vspace{-4.5mm}
\begin{itemize}[leftmargin=*]
    \item Flip the antenna to match the cartography
    \item Clean from time to time
\end{itemize}

\verspb \hline \verspt

Signal amplifiers &
\vspace{-4.5mm}
\begin{itemize}[leftmargin=*]
    \item Easy to purchase even with specific requirements
\end{itemize} &
\vspace{-4.5mm}
\begin{itemize}[leftmargin=*]
    \item Considerable noise added when increasing power
    \item Low protection against outdoor conditions
\end{itemize} &
\vspace{-4.5mm}
\begin{itemize}[leftmargin=*]
    \item Push most of the infrastructure indoor and keep just the antenna outdoors
    \item Add filters to clean and isolate the working spectrum
\end{itemize}

\verspb \hline \verspt

Signal Cables &
\vspace{-4.5mm}
\begin{itemize}[leftmargin=*]
    \item Easy to purchase and connect
\end{itemize} &
\vspace{-4.5mm}
\begin{itemize}[leftmargin=*]
    \item Considerable losses for distant equipment
    \item Difficult to adapt the installation to construction of cableways and conduits
\end{itemize} &
\vspace{-4.5mm}
\begin{itemize}[leftmargin=*]
    \item Study the different locations of equipment and use radio elements and Base Station (BS) that are as close as possible
\end{itemize}

\verspb \hline \verspt

Xilinx ZCU111 RFSoC &
\vspace{-4.5mm}
\begin{itemize}[leftmargin=*]
    \item Wideband RF signals synthesis and sampling
    \item Adaptive digital signal processing techniques
\end{itemize} &
\vspace{-4.5mm}
\begin{itemize}[leftmargin=*]
    \item SMA screw-on connectors in the XM500 daughterboard may become loose, inducing large losses
    \item Overflows in adaptive systems may render the system permanently inoperable
\end{itemize} &
\vspace{-4.5mm}
\begin{itemize}[leftmargin=*]
    \item Apply epoxy-based glue to screws 
    \item Maintain sufficient guard bits in quantized signals 
\end{itemize}

\verspb \hline \verspt

Fiber Switch &
\vspace{-4.5mm}
\begin{itemize}[leftmargin=*]
    \item High performance with low latency and high capacity
\end{itemize} &
\vspace{-4.5mm}
\begin{itemize}[leftmargin=*]
    \item Difficult to apply VPN/VLAN configuration
\end{itemize} &
\vspace{-4.5mm}
\begin{itemize}[leftmargin=*]
    \item Add an extra router to apply firewall rules
\end{itemize}

\verspb \hline \verspt

VLAN-based Circuits &
\vspace{-4.5mm}
\begin{itemize}[leftmargin=*]
    \item Application of VLANs and interconnection is feasible
    \item L2 and L3 interconnection
\end{itemize} &
\vspace{-4.5mm}
\begin{itemize}[leftmargin=*]
    \item Deep design for infrastructure isolated from corporate systems
    \item First service establishment (i.e. no previous circuit provisioned) in the order of months. Subsequent setup is considerably faster
    \item A local operator operates each area but coordinated configuration involves the national operator
\end{itemize} &
\vspace{-4.5mm}
\begin{itemize}[leftmargin=*]
    \item Consensus on design and timings across federated infrastructures to favour multiple operators alignment
\end{itemize}

\verspb \hline \verspt

Ettus B210 SDR &
\vspace{-4.5mm}
\begin{itemize}[leftmargin=*]
    \item RAN slicing
    \item Deployment of 4G networks with OAI
\end{itemize} &
\vspace{-4.5mm}
\begin{itemize}[leftmargin=*]
    \item During the tests, the number of connected User Equipment (UE) was inconsistent, posing challenges for network monitoring due to occasional instances where more UEs were displayed as connected than the actual physical count on the network
    \item Difficult to configure with OAI
\end{itemize} &
\vspace{-4.5mm}
\begin{itemize}[leftmargin=*]
    \item In-depth tinkering of the OAI cell's configuration and parameters
    \item Moving the UEs closer to the device can sometimes improve its performance; however, other times, it can worsen performance
\end{itemize}

\verspb \hline \verspt

\caption{\label{tab:hw}List of equipment studied.}
\end{longtable}

\subsection{Federation}

When setting up the networking scheme to interconnect the different sites, following some basic guidelines can reduce troubleshooting time. The guidelines are provided below. Note that the three first steps are performed at the beginning, whilst the two latter are carried out continuously and manually.

\begin{enumerate}
\item Divide a private IPv4 range into subnetworks across federating sites, ensuring that each site receives a fair amount of the addressing space (e.g. equally splitting it).
\item Consider potential collisions with other classes of private IPv4 addresses. In this case, we opted for the 10.15.0.0/16 addressing scheme to avoid clashes with the 172.16.0.0/12 and 192.168.0.0/16 schemes, which are typically used for other internal deployments.
\item When further splitting the assigned private IP subnetworks per site, it is convenient to define consequent IPs for each cluster/server.
\item Some degree of automation and configuration of the management system is ideal to maintain the virtual nodes, considering potential events like shutdowns of the hardware due to maintenance activities or power failures. Terraform and Ansible are good candidates. At the very least, network configuration should be tracked via tools such as Netplan and any manual interface or routing configuration that will not replicate across boots should be avoided.
\item Clear and up-to-date documentation on networking must be maintained by all (or most) involved members per site. For each site, it is advisable to keep: (1) a live topology/interconnection diagram indicating clusters, addressing schemes and VLANs (as well as intermediate networking devices); and (2) a list of each deployed node along with networking data (network, IP and VLAN per interface) and the cluster it runs on, in the case of virtual nodes.
\item The application of corporate or network policies from each network administration will likely require a VPN configuration to isolate the network infrastructure from the rest of the corporate system and enables the application of specific routing rules for incoming and outgoing traffic from/to other infrastructures. In this regard, routers compatible with pfSense technology are a good option. This process is carried out manually, requiring an agreed configuration across all connected infrastructures. Thus, the routing of packets, the monitoring and control systems are allowed to act as a federated infrastructure. A detailed list of services are required to connect including the employed ports.
\end{enumerate}

\subsection{Spectrum use permission}
The regulation of unlicensed spectrum that are essential for non-public networks is heterogeneous across Europe and the rest of the world, and each country is also taking position in a different time. This makes migration of a setup or finding appropriate equipment challenging, as manufacturers and vendors focus on the bands that are widely used in some countries and that are in high demand. As a result, the equipment they provide might not operate in other bands.

The national administration supports research entities in obtaining temporal and local permissions, quickly responds to questions, and fully supervises the documentation submission and the whole procedure. To obtain the permission, a public tax, which depends on the transmitted signal power, geographic area, number of systems, and planned use corners, need to be paid. This is important, as not only the number of cells, location and setup have to be declared, but also the User Equipments (UEs) that would use the radio infrastructure. This means that an study and estimation of the full communications scale is needed. 

Furthermore, it is necessary to consider two core aspects. Firstly, it is important to emphasize that the process of putting a network into operation, whether it is for indoor or outdoor setups, involves several distinct steps, such as obtaining spectrum permission from the national administration and coordinating with local administrations during the operational setup. The presence of sensible activities and population centres, such as military, health or education centres, will limit the eligible area to be covered or at least require a further radiation study to be conducted. Second, it is important to note that the permission granted is temporary and cannot be renewed. However, new permission can be requested periodically. This means that depending on the applicable regulation, the planned auctions, and the concurrency of requests for a specific area, the granted permission could change or be interrupted. As a result, equipment in a setup that works in a particular frequency band may not be operational or usable in a new one.

% \subsection{Security}

% As a neglected aspect during testbed deployments, security plays a special role in federations; specially if considering open up to the general public.

% \begin{enumerate}
% \item Network isolation must be pursued across sites from the design of the network connectivity.
% \item When federating sites at L2 or L3 layers, authentication is missing. Proper authentication schemes shall be considered either at each node open to other federated parties or at a primary gateway or bastion set to that effect.
% \item When opening the testbed to the general public, proper access controls shall be present at the hypervisor and virtual nodes themselves.
% \end{enumerate}

\section{Challenge Forecasts}
\label{sec:forecast}
% \hl{Future Challenges \\
% Those issues that will be fixed \\
% Other issues that will remain}

This section summarises some of the open challenges of the upcoming mobile networks and their enablers, and provides a subjective prediction on some current challenges that will remain as open without an evidence of being improved or addressed at the moment. To this end, Table \ref{tab:challenges} summarises the consortium perspectives based on the activities performed with the Open-VERSO testbed.

\begin{longtable}[c]{p{0.12\textwidth}p{0.35\textwidth}p{0.43\textwidth}}

\hline \verspb

\textbf{Category} & \textbf{Likely fixed} & \textbf{Likely persistent}

\verspb \hline \verspt

Spectrum permission &
\vspace{-4.5mm}
\begin{itemize}[leftmargin=*]
    \item Regulation will be more clear
    \item Procedures will be more familiar for administrations
    \item More vendors will support unlicensed bands
    \item New vendor ecosystem and products will focus on NPNs
    \item vNPNs will make deployments affordable
\end{itemize} &
\vspace{-4.5mm}
\begin{itemize}[leftmargin=*]
    \item Lack of a spectrum for NPNs will remain
    \item Commercial networks will continue to focus on big deployments and common configurations
\end{itemize}

\verspb \hline \verspt

Equipment &
\vspace{-4.5mm}
\begin{itemize}[leftmargin=*]
    \item Integrable cellular boards
    \item More competitive price
    \item Shorter delivery times
    \item Open-source drivers for the widely integrated Qualcomm SOC
    \item Wider and more nurtured SDR and SDN portfolio
\end{itemize} &
\vspace{-4.5mm}
\begin{itemize}[leftmargin=*]
    \item Not sufficiently energy efficient
    \item More competitive radio technologies for IoT
    \item Requiring an external case to make the equipment weather proof
    \item Late adopters of new 3GPP release features
    \item Limited support for specific OS versions and drivers
    \item Lack of APIs for telematic command or monitor
    \item Fragile manipulation of parts
\end{itemize}

\verspb \hline \verspt

Software &
\vspace{-4.5mm}
\begin{itemize}[leftmargin=*]
    \item Higher TRLs and granular user role control
    \item Higher stability and high-performance
    \item More APIs to achieve accurate telemetry and realize advanced remote commands
    \item Systems to reduce the complexity of configuration and operation
    \item Application of soft changes
\end{itemize} &
\vspace{-4.5mm}
\begin{itemize}[leftmargin=*]
    \item Too many options interoperable but difficult to integrate beyond families of technologies
    \item More granular life-cycle management of VNFs removing the need to reboot to update configurations
\end{itemize}

\verspb \hline \verspt

Orchestration &
\vspace{-4.5mm}
\begin{itemize}[leftmargin=*]
    \item Smooth changes and soft transitions between current and proposed configurations with penalties on configurations which can affect or interrupt the traffic at some point
    \item Control of Radio, Core and Edge resources as logical slices
    \item QoS/QoE-centric actuations evaluating metrics coming from applications, not just L1/L2/L3 metrics probed from the network nodes
    \item Normalisation of local metrics from multiple network domains to assess end-to-end values and statistics
\end{itemize} &
\vspace{-4.5mm}
\begin{itemize}[leftmargin=*]
    \item Hurdles for universal orchestration on multiple domains and multi-vendor equipment
    \item Conflicts between distributed and central controller for multiple networks with different operators
    \item Coordination of changes when involving different networks to ensure continuity with regard to connectivity
\end{itemize}

\verspb \hline \verspt

Standard &
\vspace{-4.5mm}
\begin{itemize}[leftmargin=*]
    \item IaaS vendors providing cellular stacks with industry level quality
    \item More adopters of cellular technologies for NPNs
    \item Multi-RAT infrastructures and solutions combining WiFi and Cellular
\end{itemize} &
\vspace{-4.5mm}
\begin{itemize}[leftmargin=*]
    \item Too many features not implemented in commercial or experimental setups to scale
    \item Ambition to serve a unique network for all industries and use cases without salient quotas 
    \item Slow roaming that is not transparent
\end{itemize}

\verspb \hline \verspt

\caption{\label{tab:challenges}List of Future Challenges.}
\end{longtable}

\section{Conclusions}
\label{sec:conclusions}
%\hl{Be positive}
The revolution of 5G supported by Virtualisation and Softwarisation pillars opened this industry, which was heavily dominated by big technology drivers, to newcomers and researchers, who have provided SW solutions on top of mature Big Data technologies and HW systems with standard APIs and open-source stacks with telematic and programmable interfaces. The goal is to balance the cost-performance trade-off to make networks powerful yet effective given the available resources.

Building a private 5G network for experimentation, including multiple federated infrastructures, is challenging; and requires connecting several technologies which are continuously evolving. 

There are several alternatives, equipment, software frameworks, and open-source solutions that could be adopted and integrated to create a 5G network enabling advanced and smart solutions. To fully realise 5G and 6G ambitions, it is essential to move away from a one-size-fits-all network approach and instead establish a symbiotic relationship between the network and applications' traffic.

The main goal of this document is to provide an overview of different corners to be considered, when designing, building and operating a non public network, by any institution, company or initiative in order to meet expectations and address forecast hurdles. This can be done from the experience accumulated by Vicomtech, Gradiant and i2CAT in the project Open-VERSO where a federated infrastructure for 5G/6G experimentation has been created. To this end, this document summarises existing 5G private networks for experimentation built from local, 5GPPP, or SNS initiatives. It also lists the use cases widely identified in 5G and 6G, with a focus on the ones developed in Open-VERSO to understand the context and conditions that need to be evaluated when stacks are added to built infrastructure. Then, the key identified technologies and their limitations and challenges are highlighted. Lastly, the lessons learned for different pieces composing the network and procedures to pay attention to are discussed.

This document provides a forecast on the general problems present in experimentation 5G/6G networks which will likely/ be addressed and others that, according to the historical progression on the level of maturity and the published features' roadmap, will not be solved.

This means that considerable amount of work needs to be done to meet the diverse and demanding requirements posed by representative and challenging applications and services to realise smart, intelligent and proactive behaviour of management systems. This system is expected to embed holistic understanding of the current and incoming traffic, and provide a global view on problems. Moreover, it should have the ability to locally fix or prevent them from occuring while avoiding impact on ongoing communications and applying cost-effective rules.

Further standard directions on radio and management will widen the scope to embrace even more use cases and will need more specialised solutions.

\section{Acknowledgement}
This research was supported by the Spanish Centre for the Development of Industrial Technology (CDTI) and the Ministry of Economy, Industry and Competitiveness under grant/project CER-20191015/Open, Virtualised Technology Demonstrators for Smart Networks (Open-VERSO).

\pagebreak

\section{List of Contributors}

Table \ref{tab:contribs} shows the list of contributors to this document.

\begin{table}[h]
\begin{tabular}{cl}
\hline \verspb

\textbf{Affiliation} & \multicolumn{1}{c}{\textbf{Name}}

\verspb \verspb \hline \verspt

VICOMTECH   & \begin{tabular}[c]{@{}l@{}}Mikel Zorrilla, Zaloa Fernandez, Alvaro Gabilondo, Juncal Uriol, Felipe Mogollon,\\ Mikel Serón, Michalis Dalgitsis, Roberto Viola, Angel Martin\end{tabular}

\verspb \verspb \hline \verspt

GRADIANT    & \begin{tabular}[c]{@{}l@{}}Luis Roca, Carlos Giraldo, Pablo Gonzalez, Anxo Tato, Joaquín Escudero,\\ Alvaro Vazquez, Pablo Losada\end{tabular}

\verspb \verspb \hline \verspt

i2CAT       & \begin{tabular}[c]{@{}l@{}}Daniel Camps, Andrés Cárdenas, Carlos Herranz, Joan Josep Aleixendri, Rebeca\\ Iglesias, Gianluca Cernigliaro, Mario Montagud, Pau Tomàs, Sergio Giménez,\\ Carolina Fernández\end{tabular}

\verspb \verspb \hline

\end{tabular}
\caption{\label{tab:contribs}List of Contributors.}
\end{table}

% \section{Backup}

% To view tutorials, user guides, and further documentation, please visit our \href{https://www.overleaf.com/learn}{help library}, or head to our plans page to \href{https://www.overleaf.com/user/subscription/plans}{choose your plan}.

% See the code for Figure \ref{fig:frog} in this section for an example.

% \begin{figure}
% \centering
% \includegraphics[width=0.3\textwidth]{frog.jpg}
% \caption{\label{fig:frog}This frog was uploaded via the file-tree menu.}
% \end{figure}

% You can make lists with automatic numbering \dots

% \begin{enumerate}
% \item Like this,
% \item and like this.
% \end{enumerate}
% \dots or bullet points \dots
% \begin{itemize}
% \item Like this,
% \item and like this.
% \end{itemize}

% Let $X_1, X_2, \ldots, X_n$ be a sequence of independent and identically distributed random variables with $\text{E}[X_i] = \mu$ and $\text{Var}[X_i] = \sigma^2 < \infty$, and let
% \[S_n = \frac{X_1 + X_2 + \cdots + X_n}{n}
%       = \frac{1}{n}\sum_{i}^{n} X_i\]
% denote their mean. Then as $n$ approaches infinity, the random variables $\sqrt{n}(S_n - \mu)$ converge in distribution to a normal $\mathcal{N}(0, \sigma^2)$.

% You can then cite entries from it, like this: \cite{greenwade93}. Just remember to specify a bibliography style, as well as the filename of the \verb|.bib|. Contact form at \url{https://www.overleaf.com/contact}.

\pagebreak

\bibliographystyle{alpha}
\bibliography{main}

\end{document}